\documentclass[reprint,prx,aps,longbibliography,footinbib,floatfix,nobalancelastpage,superscriptaddress]{revtex4-2}

\usepackage{multirow}
\usepackage{amsfonts}
\usepackage{amssymb}
\usepackage{mathtools}
\usepackage{graphicx}
\usepackage{subfigure}
\usepackage[dvipsnames]{xcolor}
\usepackage[colorlinks=True,citecolor=blue,linkcolor=blue,urlcolor=blue]{hyperref}
\usepackage{hypcap}
\usepackage{yfonts}
\usepackage{bm}
\usepackage{soul}
\usepackage{ulem} 
\usepackage{dsfont}
\usepackage{braket}

\usepackage{color,xcolor, soul}

    \definecolor{darkblue}{rgb}{0,0,.65}
    \definecolor{darkgreen}{rgb}{0.3,0.9,0.3}
    \definecolor{darkorange}{rgb}{0.85,0.65,0.3}
    \definecolor{cyan1}{rgb}{0.0, 0.6, 0.6}

\definecolor{myBlue}{RGB}{31,119,180}
\definecolor{myOrange}{RGB}{255,127,14}
\definecolor{myGreen}{RGB}{44,160,44}
\definecolor{myRed}{RGB}{214,39,40}
\definecolor{myPurple}{RGB}{148,103,189}

\makeatletter
\def\p@figure{\color{myBlue}}
\def\p@equation{\color{myRed}}
\makeatother

\begin{document}

\title{Semi-localized ground state in a 1D system with long-range hopping}
\author{Murod S. Bahovadinov}
\email{m.bahovadinov@rqc.ru}
\affiliation{Russian Quantum Center, Skolkovo, Moscow 121205, Russia}
\affiliation{Laboratory for Condensed Matter Physics, National Research University Higher School of Economics, Moscow, 101000, Russia}

\author{Faridun N. Jalolov}
 
\affiliation{Russian Quantum Center, Skolkovo, Moscow 121205, Russia}
\affiliation{Skolkovo Institute of Science and Technology, Bolshoy Boulevard 30, bld. 1, Moscow 121205, Russia}

\author{Vladimir E. Kravtsov}
\affiliation{ICTP, Strada Costiera 11, 34151, Trieste, Italy}

\author{Boris L. Altshuler}
\affiliation{Physics Department, Columbia University, 538 West 120th Street, New York, New York 10027, USA}
\author{Georgy V. Shlyapnikov}
\affiliation{Russian Quantum Center, Skolkovo, Moscow 121205, Russia}
\affiliation{Université Paris-Saclay, CNRS, LPTMS, 91405 Orsay, France}
\affiliation{Van der Waals-Zeeman Institute, Institute of Physics, University of Amsterdam, Science Park 904, 1098 XH Amsterdam, The Netherlands}

\begin{abstract}
We consider localization of a quantum particle with hopping amplitudes $t(r) \propto r^{-a}$ in $1D$ in the presence of diagonal disorder. Unlike the standard one-dimensional Anderson model ($a$ → $\infty$), where all states are localized, and the localization length is minimal at the band edge, in the present model for 1 $< a <$ 3/2 one has a disorder-driven localization transition at the band edge whereas high-energy states remain localized at any disorder strength. We investigate this transition for the ground state in the momentum space. We obtain perturbative expressions for the characteristic functions and moments of the wavefunction, as well as the fractal dimensions in the weak-disorder regime in the momentum space. It is demonstrated that the ground state of this model exhibits {\it semi-localization} rather than genuine localization, thus extending the list of models demonstrating an unusual {\it semi-fractality} of wavefunctions.

\end{abstract}

\maketitle

    {\it Introduction.} Anderson localization in one-dimensional disordered lattices is a paradigmatic phenomenon in which single-particle eigenstates become exponentially localized due to destructive quantum interference induced by disorder \cite{PhysRev.109.1492}. The presence of long-range hopping qualitatively modifies this picture: when hopping amplitudes decay algebraically as \(t(r)\sim 1/r^a\), tunneling between distant sites competes with disorder and can substantially alter localization properties, producing modified localization lengths, mobility edges, and unconventional spectral statistics \cite{malyshevmain, PhysRevB.99.104203, PhysRevLett.126.153201, PhysRevB.100.174201, PhysRevB.81.125104, PhysRevB.102.174203}. Such long-range disordered systems are present naturally in cold-atom, photonic, and electronic platforms with tunable non-local couplings \cite{review-Ruffo}.  

A particular realization of this physics is provided by the one-dimensional Anderson model with deterministic power-law hopping, $t_{ij}\propto \frac{1}{|i-j|^a}$,
\cite{10.1119/1.1593660, MALYSHEV2004269, PhysRevB.70.172202}. Previous works established that whereas bulk eigenstates remain power-law localized in the real space even for very slowly decaying hopping \cite{PhysRevLett.120.110602, PhysRevB.99.104203}, the states at the spectral edge undergo a disorder-driven localization-delocalization transition and are delocalized at \(1<a<3/2\) \cite{malyshevmain}. For \(a>3/2\), all states remain localized in real space for arbitrary disorder strength. Despite extensive studies of localization and mobility-edge formation in this model, detailed physical properties of the spectral-edge states remain poorly understood. To the best of our knowledge, momentum space wavefunctions were not considered previously. 

The problem of eigenfunction statistics in the {\it momentum space} of the disordered long-range hopping is the main focus of the present work.
\begin{figure}[h]
\includegraphics[width=  \columnwidth]{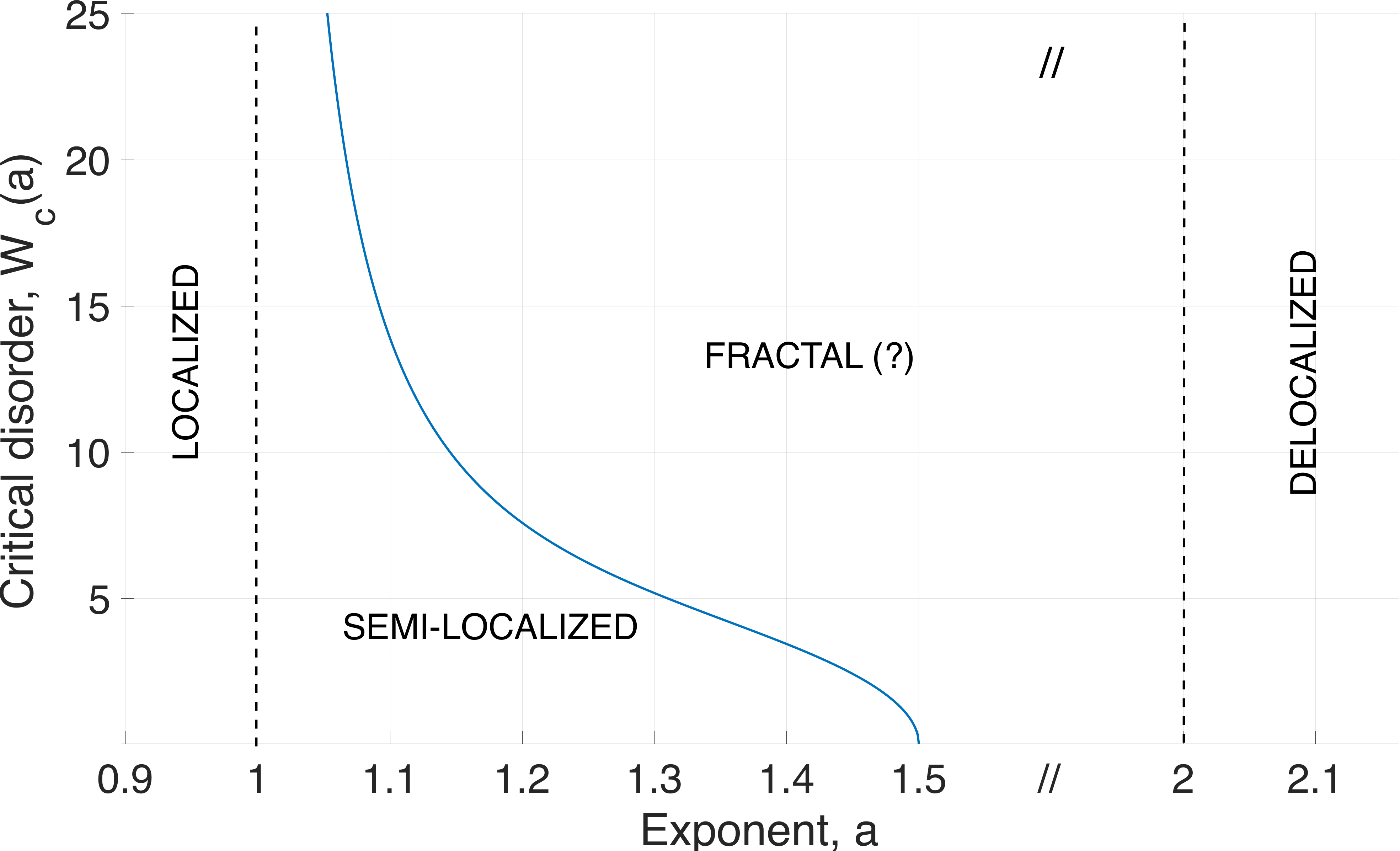}
\caption{Sketch of the phase diagram of the ground state in the momentum space. }
\label{fig:phase_dia}
\end{figure}
 
 At the same time, a number of works, recent and earlier \cite{SS2013,Ossipov2016,PhysRevB.110.174202,temkin, bnr3-5dcw,
Google_expt,VKA}, have revealed that a broad variety of disordered quantum systems exhibit a peculiar fractal structure of eigenstates characterized by generalized fractal dimensions of the form
\begin{equation}
D_q=1 \quad \text{for } q<q^*,
\qquad
D_q<1 \quad \text{for } q>q^*>1.
\end{equation}
which we refer to as  {\it semi-fractality} \cite{VKA}. 
Such states exhibit  multifractal properties only for $q>q^{*}>1$, whereas the support set fractal dimension is $D_{1}=1$.

In particular, in a recent work \cite{temkin}   the Malyshev problem \cite{10.1119/1.1593660, MALYSHEV2004269, PhysRevB.70.172202} was considered in higher dimensions  $d>1$, and it was shown that it exhibits a semi-fractal statistics of the  edge $E=0$ states in the {\it real space}. In one dimensions the $E=0$ states in the real space are completely delocalized. We believe that this result is due to the {\it hidden chiral symmetry} at the   band edge as it is shown in this paper.

In certain limiting cases one finds the most singular behavior
\begin{equation}
D_q=1 \quad (q<1),
\qquad
D_q=0 \quad (q>1).
\end{equation}
 By analogy with semi-fractality we refer to this state, which is a  hybrid of ergodic and localized features,  as {\it semi-localized} state. It was first suggested in Ref.\cite{SS2013}. 
 
 The phenomenology of the semifractal state has been observed in seemingly unrelated settings, including the \(\beta\)-ensemble \cite{bnr3-5dcw}, the Anderson model on the  Erdős–Rényi graphs \cite{PhysRevB.110.174202} and  in the deformed Wigner-Dyson ensemble of random matrices \cite{Ossipov2016}.

 However, the most important example of semi-fractal behavior is experimentally observed   in programmable quantum simulators. In particular, recent experiments by the Google quantum AI \cite{Google_expt} reported   the semi-fractal statistics of the Hilbert space eigenfunction coefficients $|\psi_{n}(r)|^{2}$ in disordered interacting qubit networks.  The peculiarity of this behavior is that the distribution of logarithm of such coefficients at sufficiently strong disorder is  a perfect power-law in $|\psi_{n}(r)|^{2}$  with the power greater than 1. Such a behavior is present in the Anderson model on Random Regular Graphs (RRG) \cite{PhysRevLett.113.046806} but only in the localized phase where this power is smaller than $1/2$.

In this work, we demonstrate that the ground state of the {\it one-dimensional} disordered  model with power-law hopping provides  a simple and analytically tractable realization of {\it semi-fractality} in its extreme form of {\it semi-localization}. 
 
Focusing on the regime \(1<a<3/2\), we show that the  ground state of this model at weak disorder exhibits the semi-localized structure {\it in the momentum space}, with the generalized fractal dimensions
 
\begin{equation}
D_q=
\begin{cases}
1, & q<1,\\
D_1\ll 1, & q=1,\\
0, & q>1,
\end{cases}\label{D_q}
\end{equation}

Our analysis is based on momentum-space perturbation theory for the long-range hopping problem at weak disorder, although as we show the conclusions are valid also in the non-perturbative regime. We show that the disorder-averaged momentum-space profile \(\langle |\psi(k)|^2\rangle\) develops a finite peak (the "condensate") of order unity at \(k=0\) typical of localized states, while the perturbative tails decay algebraically with increasing momentum and give a finite contribution  to the normalization integral. This structure immediately implies semi-localized scaling of wavefunction moments and provides a transparent microscopic mechanism for the emergence of such behavior.

Beyond the specific model considered here, we argue that this mechanism is generic for a broad class of semi-fractal states: semi-fractal scaling of moments follows naturally when the ordered (sorted) amplitudes of wavefunction coefficients decay (after averaging over disorder realizations) as a power-law ( see e.g. Ref. \cite{Ossipov2016,VKA}). If, in addition, there is a finite number of the coefficients with the largest amplitudes $O(1)$ on  top of the power-law background, the semi-localized state is realized. The difference between this work and other cases of semi-fractality (see e.g. \cite{VKA}) is that the ordering of amplitudes in our problem arises naturally from the power-law hopping in real space, while in other problems with semi-fractality it should be done by sorting. The present model therefore provides an analytically controlled single-particle setting that captures the essential physics of a phenomenon observed in much more complex disordered quantum systems. Our findings are summarized in Figure~\ref{fig:phase_dia}.
\begin{figure}[h]
\includegraphics[width=0.9\columnwidth]{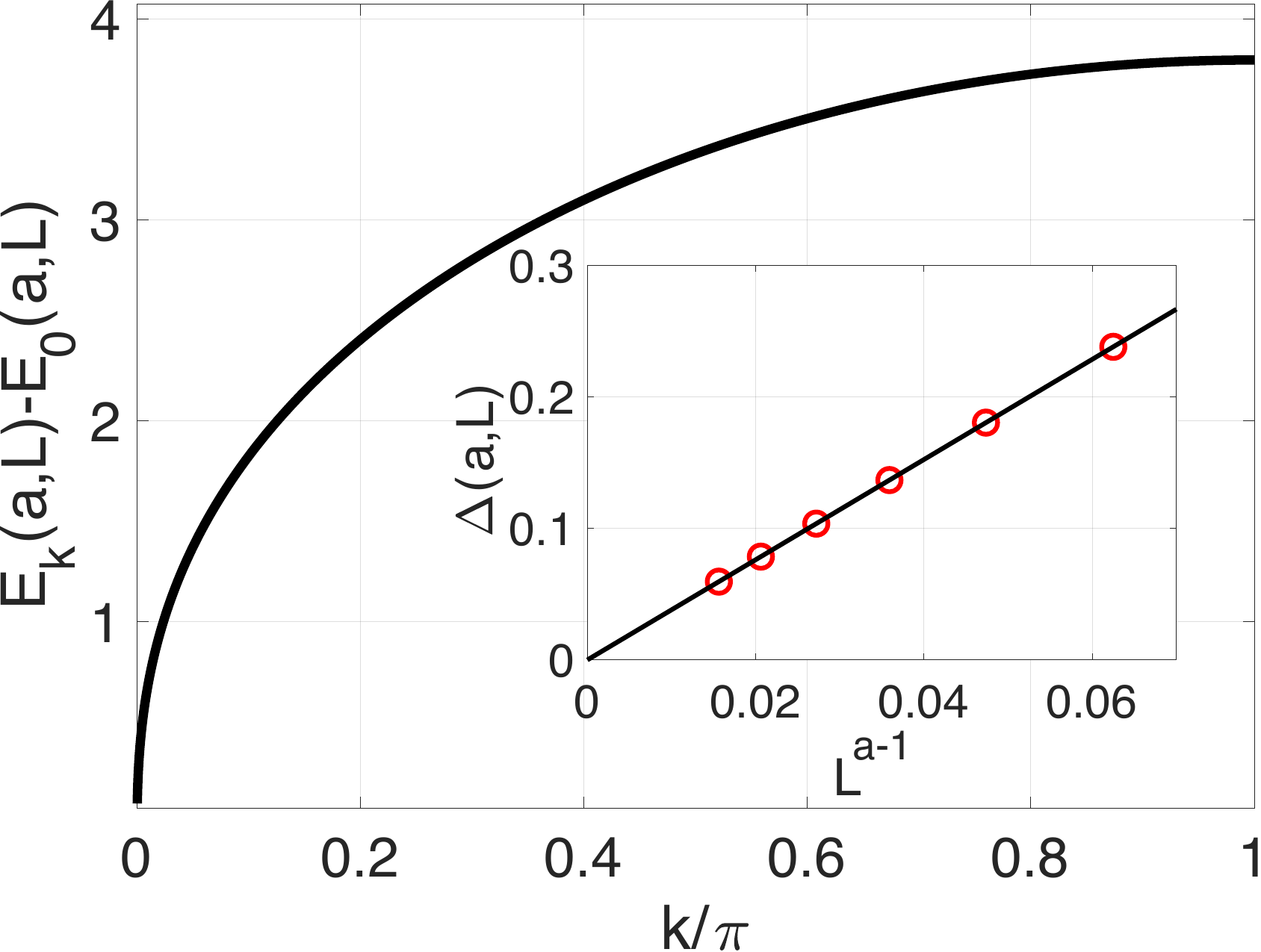}
\caption{Typical single-particle energy spectrum $E_k(a,L)-E_{0}(a,L)$ for $a=1.4$. Inset: The scaling of energy gap $\Delta(a,L)$ between the ground state and the first excited state as a function of $L^{a-1}$. The straight line is a linear fit.}
\label{fig:E_k}
\end{figure}

\textit{Model.} We consider a single particle with long-range algebraic hopping in the one-dimensional {\it real} space in the presence of on-site disorder, described by the Hamiltonian
\begin{equation}
\begin{aligned}
H &= H_D+H_t,\\
H_D &= \sum_{i=1}^{L} \epsilon_i \hat n_i,\\
H_t &= - \sum_{i\neq j}\frac{J}{|i-j|^{a}}
\left(c_i^\dagger c_j+\mathrm{H.c.}\right),
\end{aligned}
\label{eq:Hamiltonian}
\end{equation}
where \(c_i^{(\dagger)}\) annihilates (creates) a  particle in site \(i\), and we put $J=1/2$ in the power-law hopping strength.   The on-site disorder is taken to be uniformly distributed,
\begin{equation}
\epsilon_i\sim U(-W,W),
\end{equation}
with uncorrelated $\epsilon_i$ on different sites,
\begin{equation}
\langle\epsilon_i\epsilon_j \rangle
=
\frac{W^2}{3}\delta_{ij}.
\end{equation}

At weak disorder, perturbation theory is most naturally formulated in the momentum basis, where the hopping Hamiltonian is diagonal. The corresponding single-particle dispersion is
\begin{align}
E_k(a,L)= -\Big[
&\mathrm{Li}_a(e^{ik})
+\mathrm{Li}_a(e^{-ik}) \notag\\
&-\Phi(e^{ik},a,L)
-\Phi(e^{-ik},a,L)
\Big],
\label{eq:Dispersion}
\end{align}
where \(\mathrm{Li}_a(x)\) and \(\Phi(x,y,z)\) denote the polylogarithmic  and Lerch transcendent functions, respectively. The clean ground-state energy is obtained as $E_0(a,L)= -\zeta(a)+\zeta(a,L)$,
where $\zeta(a)$ is the Riemann and  \(\zeta(a,L)\) is the Hurwitz zeta functions. Notably, the explicit \(L\)-dependence of \(E_0\) implies an anomalous system-size dependence of the bandwidth, one of the characteristic features of the long-range hopping model.

In the long-wavelength limit \(k\ll1\), the dispersion admits the expansion
\begin{equation}
\begin{aligned}
E_k(a,L)=
-\Big[
&\zeta(a)
+\Gamma(1-a)\sin\!\left(\frac{\pi a}{2}\right)k^{a-1}\\
&-\frac12\zeta(a-2)k^2
-\frac1{24}\zeta(a-4)k^4
\Big]
+O(k^6),
\end{aligned}
\label{eq:LowK}
\end{equation}
valid for \(k>0\), \(1<a<3\), and \(a\neq2\). Thus, the low-energy physics is governed by a nonanalytic power-law dispersion with exponent \(a-1\). Typical energy spectrum of this form is shown in Figure~\ref{fig:E_k} for $a=1.4.$ 

Near the Brillouin-zone edge, \(k\approx\pi\), the spectrum becomes quadratic,
\begin{equation}
\begin{aligned}
E_k(a,L)\sim
&\,\eta(a)
-\frac12\eta(a-2)(\pi-k)^2\\
&+\frac1{24}\eta(a-4)(\pi-k)^4
+O[(\pi-k)^6],
\end{aligned}
\label{eq:HighK}
\end{equation}
where
\begin{equation}
\eta(x)=(1-2^{1-x})\zeta(x).
\end{equation}

The disorder-induced off-diagonal matrix elements in momentum space are $U_{k-k'}
=
\frac{1}{L}\sum_m e^{i(k-k')m}\epsilon_m.
$
For the ground-state wavefunction, weak disorder is expected to preserve localization in momentum space around the clean minimum at \(k=0\) when \(1<a<3/2\). The usual qualitative argument comes from the comparison of the typical disorder-induced hybridization matrix elements with the clean level spacing near the band minimum. The spacing between the ground state and the first excited momentum mode scales as $\Delta(a,L)=|E_0-E_{k_1}|
\sim
\frac{C_a}{L^{a-1}}$,
where \(C_a\) is an \(a\)-dependent constant (see inset of Figure~\ref{fig:E_k}), while the typical magnitude of the off-diagonal disorder matrix elements is $\sigma_D=
\frac{W}{\sqrt{3L}}$.
 
It is therefore natural to define the hybridization parameter
\begin{equation}
\lambda=
\frac{\sigma_D}{\Delta(a,L)}
\sim
\frac{W}{L^{3/2-a}}.
\label{eq:epsilon}
\end{equation}
This parameter controls the disorder-induced mixing of momentum-space plane waves. For $1<a<\frac32$,
 one has \(\lambda\to0\) as \(L\to\infty\), implying weak hybridization and stability of momentum-space localization. By contrast, for $a\ge\frac32$,
 the hybridization parameter grows with system size, indicating the eventual breakdown of perturbative momentum-space localization due to proliferating resonances. This qualitative criterion, originally proposed by Malyshev {\it et al.}~\cite{MALYSHEV2004269}, appears to be too rough to account for the fine effects of delocalization that lead to the semi-localized rather than to the true localizaed state. 

\textit{SUSY structure and chirality of the ground state.} 
 The missing ingredient in the above reasoning is the hidden chiral symmetry that follows from the supersymmetric (SUSY) structure of the ground state. 

 Indeed, let us represent $H-E_{0}I$, where $H$ is a Hamiltonian, $E_{0}$ is its ground state energy and $I$ is an identity operator, in a factorized form:
 \begin{equation}\label{factor}
H-E_{0}I\equiv H_{-}=A^{\dagger}A.
 \end{equation}
 Since $H-E_{0}$  must have one (and only one) zero eigenvalue, the rank of the matrix $A$ must be $N-1$.
 
 One can easily check that Eq.(\ref{factor}) is satisfied by the $N\times (N-1)$ matrix $A$ given by the following  {\it singular value decomposition}:
 \begin{equation}\label{SVD}
 A=V D^{1/2}U^{\dagger}.
 \end{equation}
 Here $D^{1/2}\geq 0$ is the diagonal $(N-1)\times N$ rectangular matrix composed of square-roots of non-zero eigenvalues $D$ of $H-E_{0}I$,   $U$ is the $N\times N$ unitary matrix of eigenvectors of $H$ and $V$ is a $(N-1)\times (N-1)$ unitary matrix which provides a "gauge" freedom for a choice of $A$.
 
 Now consider the "extended Hamiltonian":
 \begin{equation}\label{chiral-H}
{\cal H}=\left(\begin{matrix}0 & A^{\dagger}\cr A & 0 \end{matrix} \right).
 \end{equation}
 It possesses the chiral symmetry
 discovered by Gade and Wegner \cite{Wegner, Gade} which implies that there exists an operator $\Sigma$ which flips the sign of the Hamiltonian ${\cal H}$:
 \begin{equation}\label{chiral_sym}
\Sigma\, {\cal H} \,\Sigma =-{\cal H}.
 \end{equation}
 For ${\cal H}$ in Eq.(\ref{chiral-H}) the matrix $\Sigma$ is a $2\times 2$ block-diagonal matrix $\sigma_{z}$.
Now consider:
\begin{equation}
{\cal H}^{2}=\left(\begin{matrix}A^{\dagger}A & 0\cr 0 &  A A^{\dagger}\end{matrix} \right),
\end{equation}
which is a block-diagonal matrix containing the SUSY partners $H_{-}=A^{\dagger}A=H-E_{0}I$ and $H_{+}=A A^{\dagger}=VDV^{\dagger}$. Notice that $H_{+}$ is a non-trivial Hermitean matrix  determined by the unitary matrix $V$ of the decomposition Eq.(\ref{SVD}) as well as by the non-zero eigenvalues $D$ of $H-E_{0}I$.

The key property \cite{SUSY} of the SUSY structure is that the Hamiltonians $H_{-}$ and $H_{+}$ share the same set of eigenvalues $\varepsilon_{-}^{(n)}$ and $\varepsilon_{+}^{(n)}$, except for the zero-energy ground state $\psi_{0}$ of $H_{-}$:
\begin{eqnarray}\label{eigenvalues}
\varepsilon_{-}^{(n+1)}&=&E_{n+1}-E_{0}=\varepsilon_{+}^{(n)}.
\end{eqnarray} 
The corresponding eigenfunctions $u_{n}$ and $v_{n}$ are given by:
\begin{equation}\label{eigenstates}
v_{n}=A\,u_{n+1},\;\;\;u_{n+1}=A^{\dagger}v_{n}
\end{equation}
Since the eigenfunctions $u_{n}$  of $H_{-}$ are also the eigenfunctions for the  "reference" Hamiltonian $H$, knowing the solution of the latter one may find a complete set of eigenfunctions $v_{n}$ (and eigenvalues) of the new "partner" Hamiltonian $H_{+}$. This was the main application of the SUSY method in quantum mechanics.

 Here we use the SUSY structure to find the eigenfunctions and eigenvalues of the extended Hamiltonian ${\cal H}$. Because of chirality, it possesses a double set of eigenstates $\Psi=(u, \pm v)$ with eigenvalues $\pm {\cal E}$. The positive semi-definite operator ${\cal H}^{2}$ then should have the same eigenstates with doubly degenerate eigenvalues ${\cal E}^{2}>0$, except for the zero energy ground state which is not degenerate.

 From Eqs.(\ref{eigenvalues}),(\ref{eigenstates}) it follows that for $n\geq 1$:
 \begin{equation}
 \Psi_{n}=(u_{n}, \pm A\,u_{n}),
  \;\;\; {\cal E}_{n}=\pm(E_{n}-E_{0}).
 \end{equation}
 But the ground state of ${\cal H}^{2}$, which is the center-of-band state for the chiral Hamiltonian ${\cal H}$, is given by:
 \begin{equation}
\Psi_{0}=(u_{0},0),
 \end{equation}
 where $u_{0}$ is the ground state of the "reference" Hamiltonian $H$. Notice that $E=0$ state of the chiral Hamiltonian ${\cal H}$ is {\it gauge-invariant} with the ground state of the initial (reference) Hamiltonian $H$, being a sector of the $E=0$ eigenvector of ${\cal H}$.

Thus all the moments of  the ground state $|u_{0}|$ of $H$ are equal to the corresponding moments of the $E=0$ state $|\Psi_{0}|$ of the chiral Hamiltonian Eq.(\ref{chiral-H}).

This statement completes the proof of the chiral nature of the ground state of any Hamiltonian $H$. If chiral symmetry imposes special statistics of the $E=0$ eigenfunction, then these statistical peculiarities will enforce the corresponding peculiarities of the ground state of $H$.

The chiral symmetry is known \cite{Wegner, Gade} to protect the system from localization. We believe (see also Ref.\cite{VKA}) that it is this symmetry that partially invalidates the simple arguments based on Eq.(\ref{eq:epsilon}) and it is behind the semi-localized (instead of genuinely localized) ground state in the momentum space at weak disorder.

We would like to stress that the above proof is valid exactly only for the ground state of $H$. For the excited states with $N>n>0$ there is a non-zero second component $A\,u_{n}\neq 0$ of the chiral eigenvector $\Psi_{n}=(u_{n}, \pm A\,u_{n})$ that makes the correspondence between the moments of eigenvectors of $H$ and its chiral counterpart much more complicated. However, the above proof can be modified and is valid also for the upper edge of the spectrum $n=N$. It is sufficient to replace $H-E_{0}$ with $E_{N}-H$.

We also note that the SUSY structure and chirality that follows from it, is not by itself a sufficient condition for a semi-localized or a semi-fractal state. The states of a chiral Hamiltonian may still be localized \cite{VKA} or  ergodic.  One needs a sufficiently low density of states at the band-edge: 
\begin{equation}\label{DoS}
\rho\sim (E-E_{0})^{\zeta},\;\;\;\zeta=\frac{2-a}{a-1},
\end{equation}
$\zeta>1$ or $a<3/2$, in order to realize the semi-localized phase which persists as long as $\zeta$ is finite, or $a>1$.
(see Fig.\ref{fig:phase_dia}).  

The same behavior (\ref{DoS}) with $\zeta=d/2-1>1$ arises near the spectral edge in the  $d$-dimensional Anderson model  for $d>4$. Thus one may expect unconventional localization properties in high dimensional Anderson models as well \cite{Syzranov}.  It would be interesting to check the ground state of the $d$-dimensional Anderson model with $d>4$ regarding the semi-localized character.

{\it  Momentum-space perturbation theory. } We now derive the leading perturbative corrections to the momentum-space wavefunctions induced by weak disorder, $W\ll1$ (see the Supplemental Material for the derivation of the second-order corrections). In the clean system, the eigenstates are plane waves labeled by momentum $k$. For the clean ground state at $k=0$, the first-order correction reads
\begin{equation}
\ket{\Psi^{(1)}}=
\sum_{k\neq0} 
\frac{\bra{k}H_D\ket{0}}{\Delta_k}\ket{k}
=
\frac{1}{L}
\sum_{k\neq0}
\frac{1}{\Delta_k}
\sum_{m=1}^{L}
e^{ikm}\epsilon_m\,\ket{k},
\label{eq:first_order_state}
\end{equation}
where $\Delta_k=E_0-E_k$ denotes the energy gap.

Fixing momentum $k$, one may analyze the probability distribution of the corresponding wavefunction amplitude. For the real part of the first-order correction one finds
\begin{equation}
\Re\,\Psi^{(1)}(k)
=
\frac{1}{L\Delta_k}
\sum_{m=1}^{L}
\cos(km)\,\epsilon_m,
\label{eq:real_part}
\end{equation}
with an analogous expression for the imaginary part.

The problem thus reduces to determining the distribution of a sum of independent random variables with nonidentical coefficients. Since the onsite disorder potentials $\epsilon_m$ are assumed independent and uniformly distributed, the finite-$L$ distribution of $\Psi^{(1)}(k)$ is of Irwin--Hall type for both real and imaginary components. The corresponding characteristic function, e.g. for the real part, is
\begin{equation}
\Phi_{\Re\Psi^{(1)}}(k,t)
=
\prod_{m=1}^{L}
\mathrm{sinc}\!\left[
\frac{W\cos(km)t}{L\Delta_k}
\right].
\label{eq:cf_finite_L}
\end{equation}

In the thermodynamic limit this distribution converges asymptotically to a complex Gaussian distribution. The characteristic function of each component becomes
\begin{equation}
\Phi_{1}(k,t)
=
\exp\!\left(
-\frac{\sigma_1^2(k)t^2}{2}
\right),
\label{eq:cf_gaussian}
\end{equation}
with variance $\sigma_1^2(k) = \frac{W^2}{6L\,\Delta_k^2}$.
  
Second-order perturbative corrections can be incorporated straightforwardly. They preserve the Gaussian form of the distribution and renormalize its variance according to
\begin{equation}
\sigma_2^2(k)
=
\frac{W^4}{18L^2\Delta_k^2}
\left[
\sum_{\substack{p\neq0 \\ p\neq k}}
\frac{1}{\Delta_p^2}
+
\sum_{\substack{p\neq0 \\ p\neq k}}
\frac{1}{\Delta_p\,\Delta_{k-p}}
\right].
\label{eq:sigma2}
\end{equation}

This perturbative framework provides direct access to the full momentum-space statistics of the wavefunction. To characterize the physical eigenstates, one must subsequently normalize the perturbative wavefunctions. The normalization procedure, together with a detailed comparison between the analytical predictions and exact-diagonalization results, is presented in the Supplemental Material.

\textit{ Mean intensities.—} We now turn to the disorder-averaged normalized momentum-space intensities
\(
\langle \rho_k\rangle
\).
Upon normalization of the perturbative wavefunction, the momentum-space tail intensities are exponentially distributed,
\begin{equation}
\rho_k=\frac{|\Psi^{(1)}(k)|^2}{\mathcal S}
\sim
\mathrm{Exp}(\lambda_k),
\label{eq:rho_exp}
\end{equation}
  \begin{figure}[h]
   \includegraphics[width=\columnwidth]{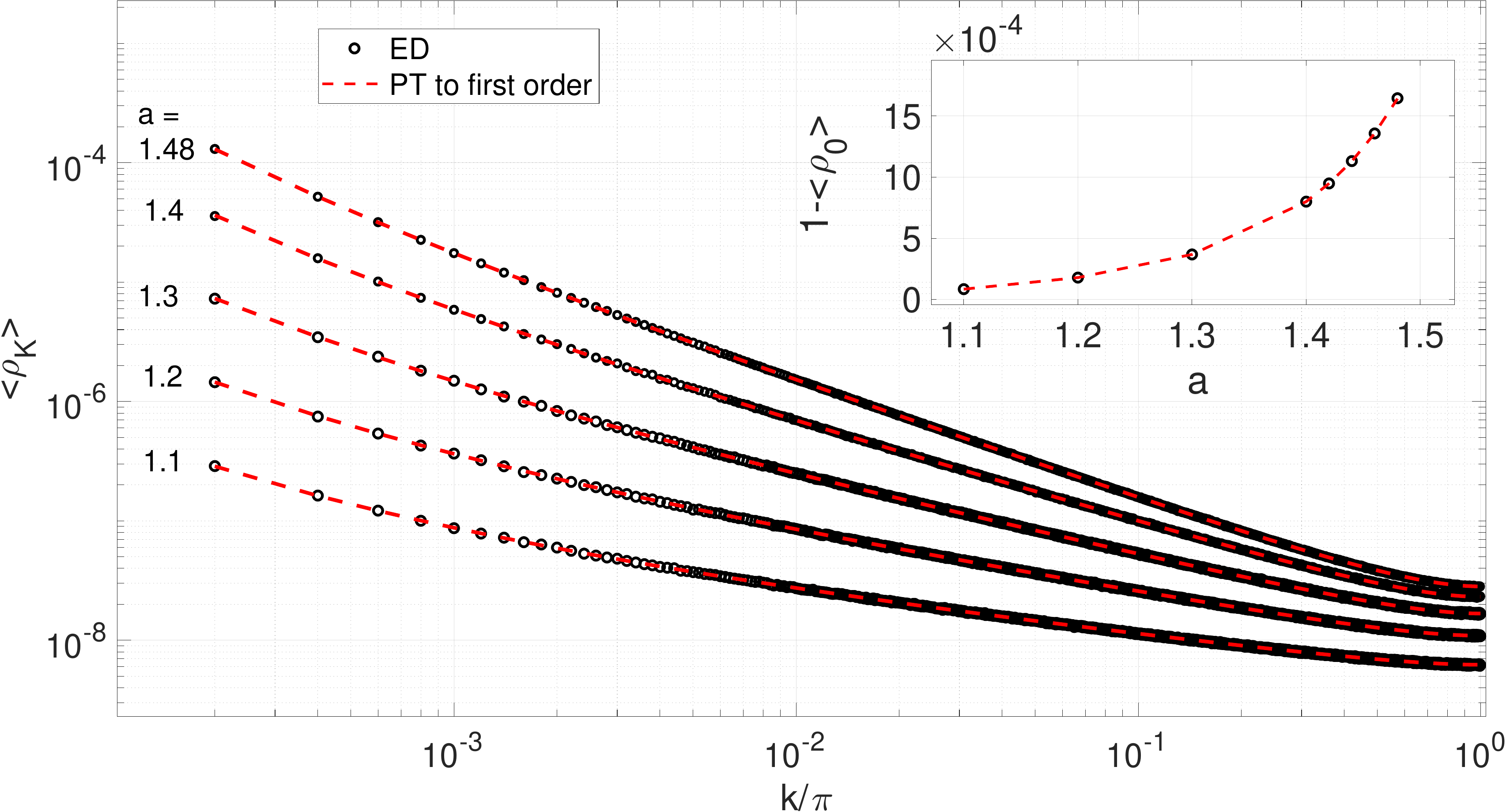}
  \caption{ED and PT results for the mean normalized tail intensities $\langle \rho_K \rangle$, shown on a log--log scale at fixed disorder amplitude $W=0.1$ for $L=10^4$. Different curves correspond to different values of $a$. Inset: $1-\langle \rho_0 \rangle$ as a function of $a$. Error bars are smaller than the symbol size. The PT results to first order reproduce the ED results with the accuracy of few percent.}
  \label{fig:meanMain}
   \end{figure}
with rate parameter at first-order,
\begin{equation}
\lambda_k=\frac{\mathcal S}{\Sigma_1^2(k)},
\qquad
\mathcal S=
1+\sum_{p\neq0}\Sigma_1^2(p),
\label{eq:lambda_def}
\end{equation}
where
\begin{equation}
\Sigma_1^2(k)=2\sigma_1^2(k)
\end{equation}
is the variance of the corresponding complex Gaussian amplitude.
 
It follows immediately that the mean normalized intensity is given by
\begin{equation}
\langle \rho_k\rangle
=
\frac{\Sigma_1^2(k)}{\mathcal S}=L^{-1}\frac{W^{2}}{3\Delta_{k}^{2}\mathcal S},
\label{eq:mean_rho}
\end{equation}
where
\begin{equation}
\Delta_{k}=E_{k}-E_{0}\propto (m/L)^{a-1},\;\;\;(m=1,2,..,L).
\end{equation}
 
Figure~\ref{fig:meanMain} compares the exact-diagonalization (ED) and perturbation-theory (PT) results for the mean normalized tail intensities \(\langle \rho_K\rangle\) at a fixed disorder strength \(W=0.1\) for $L=10^4$. We observe excellent agreement between the two approaches over the full momentum range and for all considered values of \(a\), with deviations remaining of the order  of one percent. The momentum-space tails display a clear power-law dependence inherited from the long-range dispersion relation, with an amplitude that increases systematically with \(a\).

The inset of Fig.~\ref{fig:meanMain} shows the corresponding dependence of the condensed peak depletion,
\begin{equation}
1-\langle \rho_0\rangle
=
\sum_{k\neq0}\langle \rho_k\rangle,
\end{equation}
as a function of \(a\). The depletion grows monotonically with increasing \(a\), indicating an enhanced transfer of spectral weight from the zero-momentum condensate into finite-momentum states as the dispersion becomes increasingly singular.

{\it Moments of the wavefunction and fractal dimensions.—} We now analyze the asymptotic scaling of the wavefunction moments
\begin{equation} 
P_q(L)=\sum_k \mathbb \langle \rho_k^q \rangle
=
\sum_k \frac{\langle I_k^q \rangle}{\mathcal S^q},
\label{eq:PqDef}
\end{equation}
for \(q>0\), where \(I_k\) denotes the non-normalized intensity and \(\mathcal S\) is the normalization factor. The generalized fractal dimensions are defined by
\begin{equation}
P_q(L)\sim L^{-\tau(q)},
\qquad
D_q=\frac{\tau(q)}{q-1}.
\label{eq:DqDef}
\end{equation}
The crucial point of our analysis is that the intensities $\rho_{k}$ are exponentially distributed and thus:
\begin{equation}\label{linear}
\langle \rho_{k}^{q}\rangle =\Gamma(1+q)\, [\langle \rho_{k}\rangle]^{q},
\end{equation}
where $\rho_{k>0}\propto L^{-1} |k|^{-2(a-1)}$ ($|k| L/\pi \equiv m=1, 2,...,L$) and $\rho_{k=0}\equiv \rho_{0}\propto L^{0}$ for $1<a<3/2$. After summation over $m$ in Eq.(\ref{eq:PqDef}) this results in:
\begin{equation}
P_q(L)
=
\rho_0(L)^q
+
C_2(q,W,a)L^{1-q},
\label{eq:PqScaling}
\end{equation}
and 
\begin{equation}
\tau(q)=\left\{ \begin{matrix} q-1, & 0<q<1 \cr 0, & q>1 \end{matrix} \right.,
\end{equation}
corresponding to the semi-localized $D_{q}$ of Eq.(\ref{D_q}).

The property that $\langle \rho_{k}^{q}\rangle \sim [\langle \rho_{k}\rangle]^{q}$ up to the $q$-dependent coefficient, is the basic one for the semi-fractal or semi-localized phases. The point is that it is equivalent to $\tau(q)$ which is piece-wise linear in $q$  and thus, by the Legendre transform, to the singularity spectrum $f(\alpha)$ which has a linear in $\alpha$ segment. This  is typical for the semi-localized or semi-fractal states   \cite{VKA}. 

This property is valid also for statistically homogeneous chiral systems where the moments of $|\psi(i)|^{2}$ do not depend on the site $i$ in the corresponding basis (e.g. for the eigenfunction coefficients of the Anderson model in the bulk of lattices  or the  Cayley tree \cite{VKA}). However  the averaging over disorder realizations should be performed {\it after} the sorting of the eigenfunction intensities (from largest to smallest), with the (fixed) sorting position $n$ standing for $k$ in Fig.\ref{fig:meanMain}. So, the sorted amplitudes in  statistically homogeneous chiral systems depend on a sorting position $n$ as a power-law similarly to Fig.\ref{fig:meanMain}. We believe that this power-law dependence is a common property of all chiral systems.

Now we turn to derivation of $\rho_{0}$ and $C_{2}(q,W,a)$.
We first consider the normalization factor
\begin{equation}
\mathcal S=1+\mathcal M(L),
\end{equation}
for \(1<a<3/2\). As shown in the Supplemental Material,
\begin{equation}
\mathcal M(L)
=
C_1(W,a)
\left[
\frac{1}{3-2a}
+\zeta(2a-2)L^{2a-3}
+O(L^{-1})
\right],
\label{eq:ML}
\end{equation}
where
\begin{equation}
C_1(W,a)=
\frac{2W^2}
{3\,\Gamma(1-a)^2
\sin^2\!\left(\frac{\pi a}{2}\right)
(2\pi)^{2a-2}}.
\end{equation}
Since \(2a-3<0\) in this regime, the normalization converges to a finite value in the thermodynamic limit,
\begin{equation} 
\mathcal M(L)\xrightarrow[L\to\infty]{}
\mathcal M_\infty
=
\frac{C_1(W,a)}{3-2a}.
\label{eq:Minf}
\end{equation}

with
\begin{equation}
C_2(q,W,a)
=
2\Gamma(1+q)
\left(
\frac{C_1(W,a)}
{1+\mathcal M_\infty}
\right)^q.
\end{equation}

For \(0<q<1\), the second term in Eq.~(\ref{eq:PqScaling}) dominates, yielding
\begin{equation}
\tau(q)=q-1,
\qquad
D_q=1.
\label{eq:DqLess1}
\end{equation}

By contrast, for \(q>1\), the finite condensate contribution dominates,
\begin{equation}
P_q(L)\to \rho_0(\infty)^q,
\end{equation}
implying
\begin{equation}
\tau(q)=0,
\qquad
D_q=0.
\label{eq:DqGreater1}
\end{equation}

The information dimension follows from the Shannon entropy and reads
\begin{equation}
D_1=
\frac{\mathcal M_\infty}{1+\mathcal M_\infty}.
\label{eq:D1}
\end{equation}

Collecting all contributions, the full set of fractal dimensions is
\begin{equation}
D_q=
\begin{cases}
1, & 0<q<1,\\[4pt]
\dfrac{\mathcal M_\infty}{1+\mathcal M_\infty}, & q=1,\\[10pt]
0, & q>1.
\end{cases}
\label{eq:FullDq}
\end{equation}

\begin{figure*}[t]
  \centering 
   \includegraphics[width=0.9\textwidth]{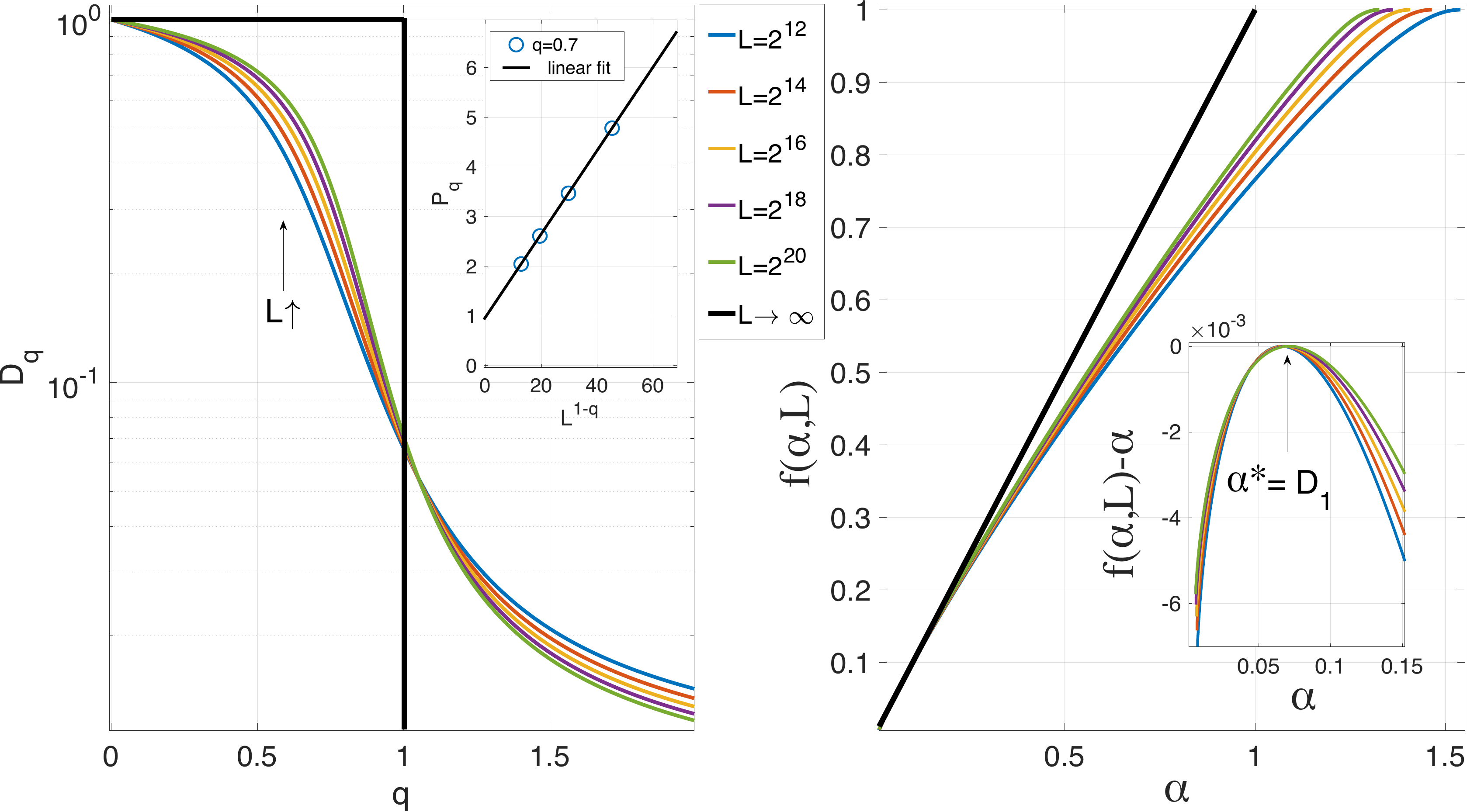}
   \caption{
Left panel: Fractal dimensions $D_q$ obtained from PT for several system sizes $L$ at fixed disorder strength $W=0.6$ and $a=1.48$. The black solid curve denotes the predicted thermodynamic-limit for $D_{q}$, Eq.~(\ref{eq:FullDq}). The finite-size data converge toward  this limit with increasing $L$ but the convergence is very slow. Inset: Finite-size scaling of $P_q$ at fixed $q=0.7$, illustrating the fit to the ansatz $P_q=a_q+b_qL^{1-q}$. The solid line is a linear fit.  
Right panel: Corresponding singularity spectrum $f(\alpha,L)$ obtained via Legendre transform of $\tau_{L}(q)=-\ln P_{q}/\ln L$. The singularity spectrum  approaches at $L\rightarrow \infty$ the semi-localized  form $f(\alpha,\infty)\equiv f(\alpha)=\alpha$, ($0<\alpha<1$), predicted by perturbation theory, which is shown by the black solid line. In the inset we plot $f(\alpha)-\alpha$ which should generically have a maximum $f(\alpha^*)-\alpha^*=0$ at $\alpha=\alpha^*=D_{1}$ \cite{SS2013}
} 
  \label{fig:DqAndFa}
   \end{figure*}

Equation~(\ref{eq:FullDq}) constitutes the central result of this work. The corresponding generalized fractal dimensions, shown in Fig.~\ref{fig:DqAndFa} (left panel), reveal a semi-localized structure in the regime \(1<a<3/2\). Representative numerical results for \(a=1.48\) and system sizes up to \(L\sim10^6\) demonstrate clear convergence towards the predicted thermodynamic-limit behavior. As the system size increases, the fractal dimensions approach \(D_q=1\) for \(q<1\), consistent with the asymptotic scaling \(P_q=a_q+b_qL^{1-q}\) used to extrapolate the moments [see inset of Fig.~\ref{fig:DqAndFa}]. Conversely, for \(q>1\), the extracted values of \(D_q\) decrease with increasing system size and tend towards \(D_q=0\) in the thermodynamic limit.
 
These results show that the eigenstates are neither fully extended nor fully localized. Rather, in the thermodynamic limit they show ergodic properties for measures that probe $q<1$ and show localized properties for probes sensitive to $q>1$. That is why they may be called {\it semi-localized} (or ${\it semi-ergodic}$ depending on the viewpoint).

The associated spectrum \(f(\alpha)\), obtained from the generalized fractal dimensions via the Legendre transform
\begin{equation}
\alpha=\frac{d\tau(q)}{dq},
\qquad
f(\alpha)=q \alpha-\tau(q),
\end{equation}
is shown in Fig.~\ref{fig:DqAndFa} (right panel).
In the thermodynamic limit it is convergent to a linear segment 
\begin{equation}
f(\alpha)=\alpha,\;\;\;(0<\alpha<1),
\end{equation}
with fixed $D_{1}$, which is the point of the maximum of $f(\alpha)-\alpha$ at a finite $L$.

For \(a\ge 3/2\), the perturbative normalization factor ceases to remain finite in the thermodynamic limit. Instead,
\begin{equation}
\mathcal M(L)\sim
\begin{cases}
\log L, & a=3/2,\\
L^{2a-3}, & 3/2<a<3,
\end{cases}
\label{eq:MLDivergence}
\end{equation}
up to higher-order perturbative corrections. Consequently,
\begin{equation}
\rho_0=
\frac{1}{1+\mathcal M(L)}
\xrightarrow[L\to\infty]{}0,
\end{equation}
and the semi-localized phase breaks down.  

Within the perturbation theory, this indicates a complete hybridization of the resonant momentum mode with the perturbative background. The wavefunction can no longer be interpreted as a finite-weight $k=0$ condensate plus weak tails. Instead, the normalization becomes dominated by the proliferating near-resonant states at small momentum, and the perturbative expansion ceases to be asymptotically controlled. At a weak disorder $W\rightarrow 0$ the point $a_c=\frac32$
 therefore marks the critical boundary separating the  semi-localized phase from the delocalized one.

 {\it  Numerical results beyond the perturbation theory. }
Within the first order perturbation theory in $W^{2}$ we obtained a power-law for $\langle |\psi(k)|^{2}\rangle$ at $kL\gg 1$:
\begin{equation}\label{obser}
\langle |\psi(k)|^{2}\rangle =C(a)\,\frac{W^{2}}{L\,k^{2a-2}}.
\end{equation}
This power-law behavior was crucial for existence of the semi-localized phase at $\mu=2a-2<1$. The question is what happens when higher-order perturbation is taken into account? Will the power-law be still maintained and if yes, will the exponent $\mu$ depend on $W$ and how? 

In this section we present the numerical results which show that the power-law holds even when $W^{2}$ is not small. In particular, we demonstrate that near $a=3/2$ the exponent $\mu$ is significantly $W$-dependent, while away from $3/2$ it is mostly unchanged by higher-orders in $W$.

\begin{figure*}
\centering{
\includegraphics[width=0.7\linewidth]{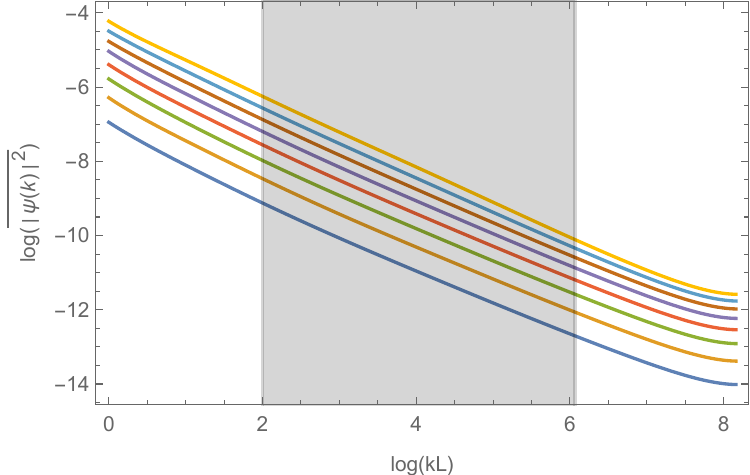} 
  
 \includegraphics[width=0.3\linewidth]{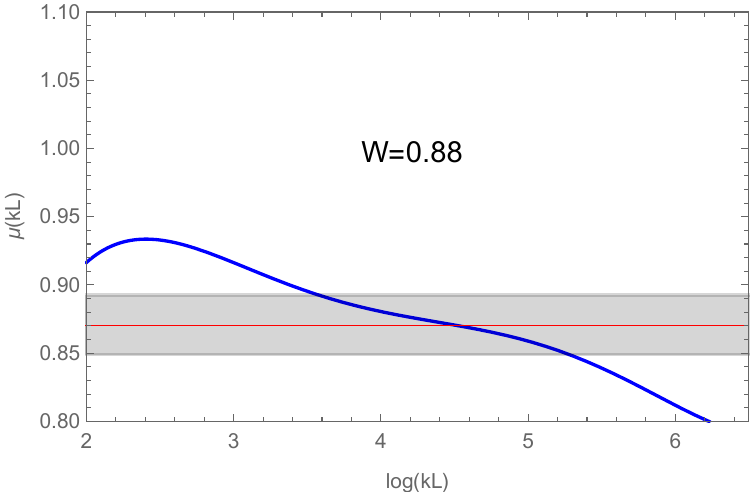}
 \includegraphics[width=0.3\linewidth]{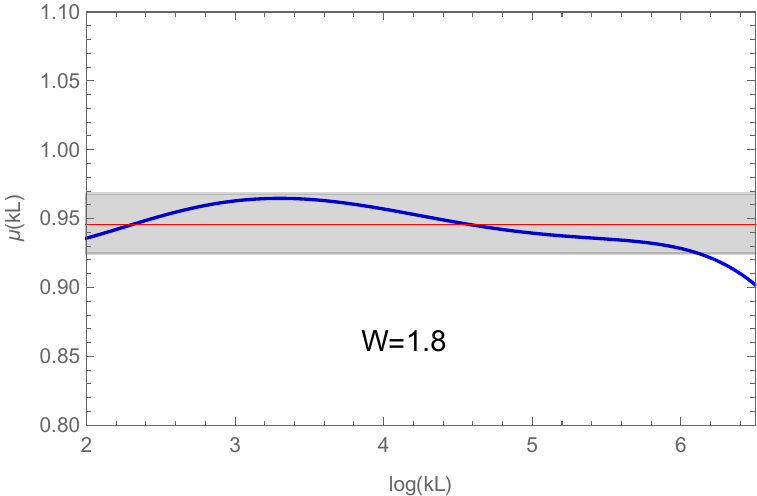}
\includegraphics[width=0.3\linewidth]{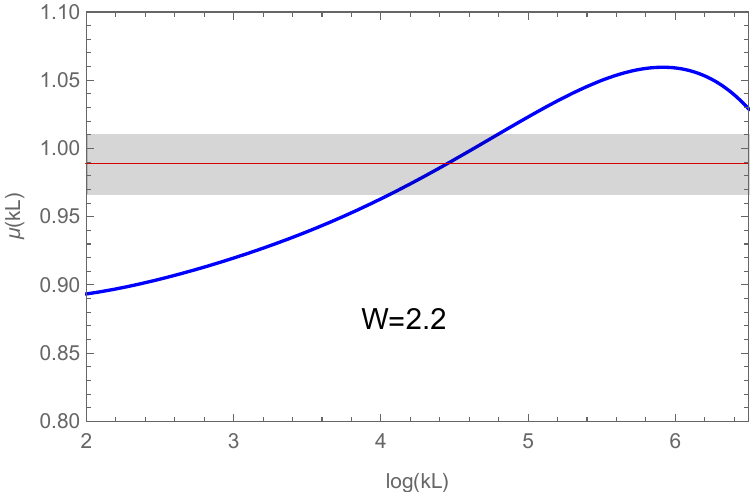}
\includegraphics[width=0.45\linewidth]{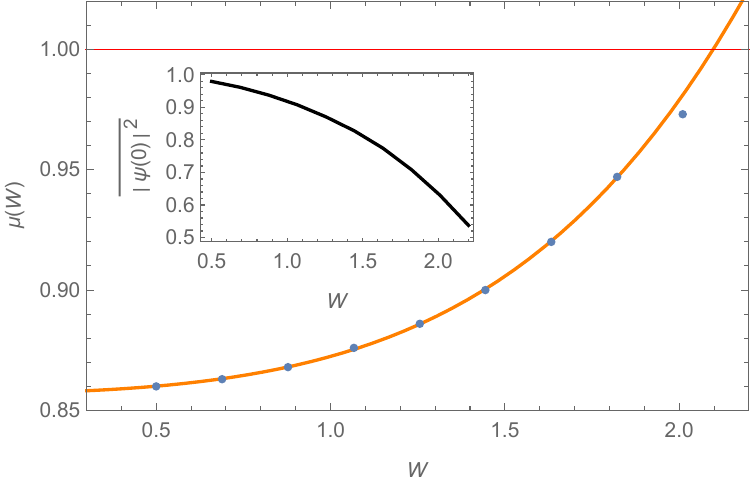}
 \includegraphics[width=0.43\linewidth]{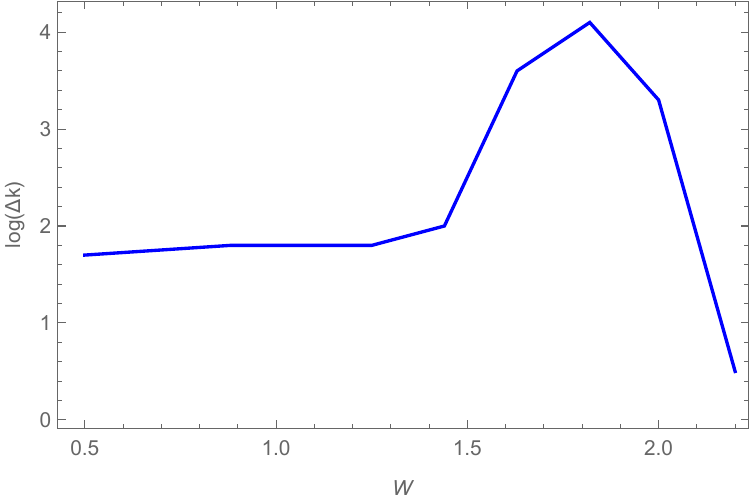}}
               
          \caption{Upper panel: $\langle|\psi(k)|^{2}\rangle$ as a function of $\ln(kL)$ for $a=1.4$ and values of disorder strength $W$ from $W=0.5$ (blue) to $W=1.8$ (yellow) computed numerically for the system length $L=7000$. Gray area is the fitting region. Middle panel: The running exponent $\mu=-d\ln\langle|\psi(k)|^{2}\rangle/d\ln(kL)$ as a function of $\ln(kL)$. The red line is the averaged $\mu(k)$ in the fitting region. The gray horizontal strip of the width $0.04$ serves to define the region of the "good power-law", i.e. the region of $kL$ where the running exponent does not deviate from the average more than by $\pm 0.02$. Lower panel: the dependence of the averaged running exponent $\mu(W)$ and the logarithm of the width $\Delta k$ of the region of "good power law" on disorder strength $W$. This region shrinks dramatically at $W\approx 2$ where the averaged running exponent reaches $\mu_{c}=\mu(W_{c})=1$.
          The inset show the behavior of $\langle |\psi(k=0)|^{2}\rangle$ as a function of $W$. At a point where $\mu(W)=1$ the amplitude of the zero mode $\langle |\psi(k=0)|^{2}\rangle\approx 0.5$ is still far from zero.  Thus the termination of the semi-localized phase is determined by $\mu(W_{c})=1$ rather than $\langle |\psi(k=0)|^{2}\rangle=0$.}
     \label{fig:a_1_p_4}
 \end{figure*}

We start with numerical results for $\langle |\psi(k)|^{2}\rangle$. First we consider the case of $a=1.4$ close to
$a=3/2$. The results of exact diagonalization for this case are shown in Fig.\ref{fig:a_1_p_4} for the values of disorder strength $W=0.5,0.69,0.88,1.07, 1.25,1.44,1.63,1.82,2.01,2.20$ for the system size $L=7000$.
This figure shows that even at $W=1.8$ the power-law holds with the accuracy $\pm 0.02$ for the exponent $\mu(k)$ in a finite interval of $k$. For very small $kL\lesssim 1$ the power-law is violated by the discreteness of $k$ in a finite system   and for $k\sim 1$ the departure from the power-law is caused by the discreteness of the lattice in the real space. Away from those two regions (see the gray area in the upper panel of Fig.\ref{fig:a_1_p_4}) the power-law is quite good. 

We fitted the curves $\langle |\psi(k)|^{2}\rangle$  for the values of $kL$ determined by $2<\ln(kL)<\ln(\pi)-2\approx 6$. The result for the averaged exponent $\mu(W)$ is shown in the lower panel of Fig.\ref{fig:a_1_p_4}. The obtained $\mu(W)$ is well approximated by $\mu(W)=0.85+0.010 W^{2}+0.005 W^{4}$ all the way from $W=0.5$ up to the point $W=W_{c}\approx 2$ where $\mu(W_{c})=1$ and the semi-localized phase must terminate. 
We further define the region $\Delta k$ of a "good power-law" by the condition that the running exponent $\mu(k)=-d\ln\langle |\psi(k)|^{2}\rangle/d\ln(k)$ is within $|\mu(k)-\mu(W)|<0.02=(3-2a)/10$. The plot of $\ln(\Delta k)$ is shown in the lower panel of Fig.\ref{fig:a_1_p_4}.
Remarkably, for $W>W_{c}$ the width of the region of a good power-law shrinks dramatically, so that it makes no sense any more to speak about the power-law behavior of $\langle |\psi(k)|^{2}\rangle$.
\begin{figure}
    \centering
    \includegraphics[width=1.0\linewidth]{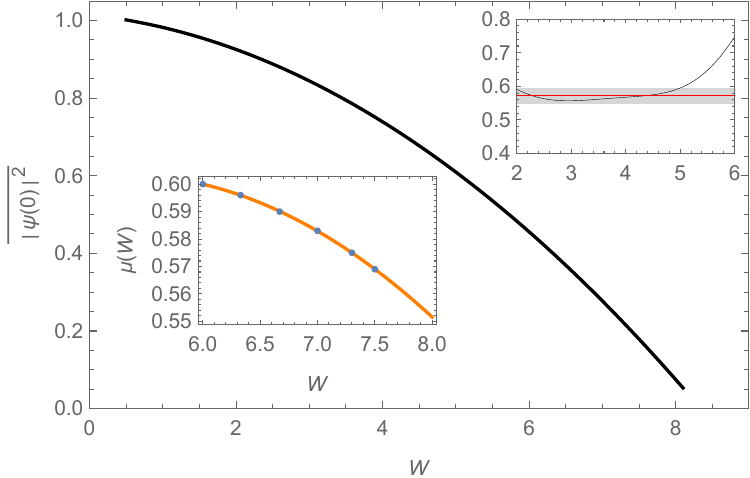}
    \caption{Main plot: The behavior of the amplitude of the zero mode $\langle|\psi(k=0)|^{2}\rangle$ on the disorder strength $W$ for $a=1.2$ obtained by exact diagonalization at $L=7000$. Lower inset: the averaged exponent $\mu(W)$. Upper inset: the running exponent $\mu(k)$ as a function of $\ln(kL)$ for $W=7.3$. In the entire region $W\in [0.5,7.3]$ the exponent $\mu(W)$ is weakly $W$-dependent varying from $\approx 0.5$ at $W=0.5$ to $\approx 0.6$ near the transition point $W_{c}\approx 8$. The transition point of termination of the semi-localized phase for $a=1.2$ is determined by vanishing of the zero mode amplitude  $\langle|\psi(k=0)|^{2}\rangle=0$ rather than by $\mu(W_{c})=1$ as it is the case for $a=1.4$.}
    \label{fig:psi_0_a_1p2}
\end{figure}

Now we report on the numerical results at $a=1.2$ far from $a=3/2$. In this case the first order perturbation theory predicts $\mu=2a-2=0.4$ at $kL\gg 1$. Exact diagonalization, however, gives a somewhat larger exponent $\mu(W)\approx 0.5$ for $2\gtrsim W\gtrsim 0.5$. However, $\mu(W)$ is very weakly dependent on the disorder strength $W$ reaching the maximum value $\mu(W)\approx 0.6$ at $W\approx 5$ and then decreasing slowly as $W$ further increases (see the lower inset in Fig.\ref{fig:psi_0_a_1p2}). Such a behavior invalidates the previous scenario (the termination of the semi-localized phase by reaching a limiting $\mu(W)=1$, realized for $a$ close 3/2). Instead, the critical disorder strength is determined by the point $W_{c}$ where the amplitude of the zero mode $\langle |\psi(k=0)|^{2} \rangle$ first vanishes. In the main plot of Fig.\ref{fig:psi_0_a_1p2} it is shown that this happens at $W\approx 8$ which is in a reasonably good agreement with the estimation:
 \begin{equation}
W_c(a)
=
\sqrt{
\frac{3}{2}
(3-2a)
\Gamma(1-a)^2
\sin^2\!\left(\frac{\pi a}{2}\right)
(2\pi)^{2a-2}
},
\label{eq :WcAlpha}
\end{equation}
based on the perturbative expression for $\mathcal M_{\infty}$, Eq.(\ref{eq:Minf}), with the condition $M_{\infty}(W_{c})\sim 1$ (see also Fig.\ref{fig:phase_dia}).

{\it Second-order perturbation theory and the guess for the RG.}
The fact that the power-law in Eq.(\ref{obser}) survives higher-order perturbation in $W^{2}$ is non-trivial. In what follows we suggest a two-parameter renormalization group (RG) at $3/2-a=\epsilon\ll 1$ which reproduces the first and second order perturbation theory and preserves the power-law, albeit with $\mu$ growing with $W$. We demonstrate that the power-law is preserved because of the $\beta$-function that is identically zero. In other problems described by the similar RG equations, the zero  $\beta$-function is related to the hidden SUSY structure. This observation makes a bridge between the power-law behavior of $\langle |\psi(k)|^{2}\rangle$ and power-law segments in the distribution function of $|\psi|^{2}$  on the one hand, and the SUSY structure and chiral symmetry of the problem on the other.

We now start with the expression for $\langle |\psi(k)|^{2}\rangle$ in the first and second order perturbation theory in $W^{2}$:
\begin{equation}\label{2_order_PT}
\langle|\psi(k)|^{2} \rangle =\frac{\pi\,\gamma_{0}}{2 L\,k^{2a-2}}\,\left(1-\gamma_{0}\,\ln k -\epsilon\,\gamma_{0}\,(\ln k)^{2}+O(\gamma_{0}^{2}) \right),
\end{equation}
where $\epsilon=0<3/2-a\ll 1$, $\epsilon\ln k\ll1, \gamma_{0}\ln k\ll1$ and 
\begin{equation}\label{gamma_0}
\gamma_{0}=\frac{W^{2}}{3\pi\,(\Gamma(1-a)\,\sin(\pi a/2))^{2}}.
\end{equation}
In deriving this expression we assumed $kL\gg 1$ and replaced the sum over $p$ in Eq.(\ref{eq:sigma2}) by the integral.

Let us introduce:
\begin{equation}\label{gamma}
\gamma(k)=\gamma_{0}\,k^{\epsilon}\,(1-\gamma_{0}\,\ln k -\epsilon\,\gamma_{0}\,(\ln k)^{2})\approx \gamma_{0}\,k^{\epsilon}\,(1-\gamma_{0}\,k^{\epsilon}\,\ln k).
\end{equation}
Then
\begin{equation}\label{psi_k-gamma}
\langle|\psi(k)|^{2} \rangle =\frac{\pi\,\gamma(k)}{2 L\,k^{2a-2+\epsilon}},
\end{equation}
It is easy to see that with the accuracy up to $\gamma_{0}\ll 1$ and the lowest order in $\epsilon$ ($0<\epsilon=3/2-a\ll 1$) we have: 
\begin{equation}\label{RG_pert}
\frac{d\ln \gamma}{d \ln k}=\epsilon-\gamma_{0}\,k^{\epsilon} + O(\gamma_{0}^{2}).
\end{equation}
Note that the true RG equations should depend only on the running parameters like $\gamma(k)$ but not on the initial conditions $\gamma(k=1)=\gamma_{0}$ or explicitly on the variable $k$.

An obvious guess for the RG equation the solution of which at $\epsilon|\ln k|\ll1$ and $\gamma_{0}|\ln k|\ll 1$ coincides with Eq.(\ref{gamma}) up to $\gamma_{0}^{2}$ order is:
\begin{equation}\label{WL_eq}
\frac{d\ln \gamma}{d\ln k}=\epsilon -\gamma.
\end{equation}
However, the full solution of this equation with the initial condition $\gamma(k=1)=\gamma_{0}$:
\begin{equation}\label{gamma_WL}
\gamma=\gamma_{0}\,k^{\epsilon}\,{\rm exp}\left[\frac{\gamma_{0}}{\epsilon}\,\left(1-k^{\epsilon} \right)\right],
\end{equation}
is not a pure power-law. Moreover, the dependence of $\langle |\psi(k)|\rangle$ on $\ln k$ is not compatible with the numerics for $\mu(k)$ at $\gamma_{0}<\epsilon$ discussed in the previous Section: $\mu(k)$ slowly increases with increasing $\ln k$ rather than to develop a plateau or to slowly decrease (see the middle panel of Fig.\ref{fig:a_1_p_4} and the upper inset of Fig.\ref{fig:psi_0_a_1p2}).

There is, however, a different choice of RG equations that makes $\gamma(k)$ a pure power-law, while reproducing Eq.(\ref{2_order_PT}). This RG is not single-parameter but rather a {\it two-parameter} one: 
\begin{eqnarray}\label{two-param_RG}
\frac{d\ln g}{d\ln k}&=&\beta(g),\\
\frac{d\ln \gamma}{d\ln k}&=&\epsilon -\zeta(g,\epsilon)\approx g+O(g^{2}).
\end{eqnarray}
In general, $\zeta(g,\epsilon)$ may also depend on $\epsilon$. Numerics suggests that this dependence is very likely of the form:
\begin{equation}
\zeta(g,\epsilon)=g\,F(g/\epsilon).
\end{equation}

The second parameter $g(k)$ allows a constant solution $g(k)=g_{0}$, provided that the $\beta$-function $\beta(g)$ is exactly zero. Then we immediately obtain a pure power-law solution for $\gamma(k)$:
\begin{equation}
\gamma(k)=\gamma_{0}\,k^{\epsilon-g_{0}},
\end{equation}
which at $\epsilon\,|\ln k|\ll1$ and $\gamma_{0}|\ln k|\ll1$ reproduces the result, Eq.(\ref{2_order_PT}), of the perturbation theory, if we choose $g_{0}=\gamma_{0}$.

Notice that $\gamma(k)$ remains a pure power-law for any $\zeta(g,\epsilon)$ provided that 
\begin{equation}\label{zero-beta}
\beta(g)\equiv 0.
\end{equation}
If this is the case, then we obtain from Eq.(\ref{psi_k-gamma}):
\begin{equation}
\langle|\psi(k)|^{2} \rangle =\frac{\pi\,\gamma_{0}}{2 L\,k^{\mu}},
\end{equation}
where
\begin{equation}\label{mu-W}
\mu\equiv\mu(W)=2a-2+\zeta(\gamma_{0},\epsilon).
\end{equation}
Since $\gamma_{0}$, Eq.(\ref{gamma_0}), is proportional to $W^{2}$, Eq.(\ref{mu-W}) determines the dependence of the exponent $\mu$ on the disorder strength $W$. In the case where the termination of the semi-localized state is due to $\mu$ reaching the limiting value $\mu(W_{c})=1$, Eq.(\ref{mu-W}) provides an equation for the critical disorder strength:
\begin{equation}
\zeta(\gamma_{0},\epsilon)=2\epsilon.
\end{equation}

In particular,  when approaching $a-3/2$ from below, one finds using $\zeta(\gamma_{0},\epsilon)\approx \gamma_{0}\propto W^{2}$:
\begin{equation}
W_c(a)\propto {(3-2a)^{\delta}}.
\qquad
\delta=\frac12,
\qquad
a\to\frac32^-,
\end{equation}
 
On the other hand, for $a \rightarrow1^+$ the termination of the semi-localized state is due to the vanishing amplitude $\langle|\psi(k=0)|^{2} \rangle$. In this case the critical disorder strength should be found from Eq.(\ref{eq :WcAlpha}), which gives:
\begin{equation}
 W_c(a)\propto {(a-1})^\delta,
\qquad
\delta=-1,
\qquad
a\to1^+,
\end{equation}

{\it Conclusion and discussion.—}
In this work, we developed a momentum-space perturbation theory for the problem of a single quantum  particle with long-range deterministic algebraic hopping $t\sim r^{-a}$ in a one-dimensional real space in the presence of weak onsite disorder of the strength $W\ll 1$. Exploiting the weak hybridization of momentum modes in the regime \(1<a<3/2\), we derived the full eigenfunction statistics  in the momentum space of the disordered ground-state wavefunction.

Our analysis demonstrates that the eigenfunction statistics in the thermodynamic limit does not match neither fully ergodic nor algebraically localized ones. It is a very peculiar multifractal statistics with the information fractal dimension that determines the volume of the eigenfunction support set, $D_{1}\propto W^{2}\ll 1$, and a localized behavior $D_{q}=0$ for $q>1$. However, for $q<1$ the behavior is analogous to that for an ergodic system: $D_{q<1}=1$. We call this type of statistics "semi-localized". 

It is the limiting case of the so called semi-fractal statistics that has been recently found in many seemingly different systems \cite{SS2013,Ossipov2016,PhysRevB.110.174202,bnr3-5dcw,
Google_expt,VKA}.  

Using the SUSY stricture at the band edge, we show that the eigenfunction statistics in the ground state  of {\it any} Hamiltonian with the finite spectral bandwidth is equivalent to the one of the center-of-band state of a manifest chiral Hamiltonian.  

We believe that in the {\it one-dimensional} Malyshev problem \cite{10.1119/1.1593660,MALYSHEV2004269, https://doi.org/10.1002/pssb.200304868,malyshevmain,PhysRevB.70.172202} at weak disorder and equivalent problems \cite{Gurarie,Syzranov} with the dispersion of a free particle $k^{a-1}$ in a clean system, the chiral symmetry leads to a semi-localized rather than to the genuinely localized state at $1<a<3/2$, despite the hybridization parameter $\lambda$ (\ref{eq:epsilon}), vanishes in the thermodynamic limit.

The point is that the semi-fractal and semi-localized eigenfunction statistics follows from the power-law behavior of the averaged sorted amplitudes $\langle |\psi(n)|^{2}\rangle$ on the sorting index $n$ \cite{VKA}. In the current problem the momentum $k$ plays the role of the sorting index. We have demonstrated that this power-law behavior follows from the perturbation theory (PT) in the disorder strength $W$ and persists in numerics well beyond the PT. We suggested the RG equations that reproduce the results of PT in the first and second order in $W$ and lead to a pure power-law behavior of $\langle |\psi(k)|^{2}\rangle$ also beyond the PT. These are quite general {\it two-parameter} RG equations, one of which must have a zero $\beta$-function in order for the pure power-law to be a solution.  A similar system of RG equations with zero $\beta$-function appeared in different problems, most notably in Ref.\cite{KLY1,KLY2,KLY3}, where a random walk in a random potential field was considered. Exactly zero $\beta$ function (proven in Refs,\cite{KLY3,Honkonen}) in this problem led to a sub-diffusion $\langle r^{2}\rangle \propto t^{\nu}$ with the 
exponent $\nu$ depending continuously on the strength of the random potential field. 

Remarkable fact is that this problem, too, has a {\it hidden SUSY structure} (see Chapter S8 in the SM) which is, therefore, a necessary ingredient for the zero $\beta$-function and the power-law behavior of  observables  that follows from it. This analogy makes us confident that, quite generically, the semi-fractal and semi-localized eigenfunction statistics emerges when the problem has an explicit or hidden SUSY structure and chiral symmetry and that the RG framework outlined in the previous Section is a proper generic description.

We estimated the critical disorder $W_{c}(a)$ of the transition between the semi-localized and extended phase and sketched the corresponding phase diagram in the momentum space, Fig.\ref{fig:phase_dia}.

What happens for $W>W_{c}(a)$ is  currently under discussion. A limited information that can be obtained from the exact diagonalization numerics shows that:
\begin{equation}
\langle |\psi(k)|^{2}\rangle \sim {\rm exp[-(m/R)^{x}]},
\end{equation}
where $m=k L/\pi$ and $R\sim L^{y}$ is an extensive parameter that diverges in the thermodynamic limit. However, for $a<2$ at weak disorder one has $y<1$. At the same time at $a=3/2$ the exponent $y$ must be zero. So, we conjecture that
\begin{equation}
y=2a-3.
\end{equation}
If this conjecture is correct then (at least at weak disorder) the phase beyond $W_{c}(a)$ and at $a<2$   is formally fractal, non-ergodic extended, with the fractal dimension $y<1$ (see Fig.\ref{fig:phase_dia} ). According to this conjecture the true ergodic phase in the momentum space (localized phase in the real space) emerges (at weak disorder) only for $a>2$. This sheds some light on the controversial results of Ref.\cite{Seperina} where the true localized behavior at the spectral edge in the real space was observed only at $a>2$.

\begin{acknowledgements}
{\it Acknowledgments.—}
We would like to thank V. Gurarie, I.M. Khaymovich and C. Vanoni for stimulating discussions.
This work was financially supported by the Russian Science Foundation (project No. 25-72-00135). 
This research was also supported in part through computational resources of the HPC facilities at HSE University. 
\end{acknowledgements}

\bibliography{Lib}


\newpage

\setcounter{equation}{0}
\setcounter{figure}{0}
\setcounter{page}{1}
\renewcommand{\theequation}{S\arabic{equation}}
\renewcommand{\thefigure}{S\arabic{figure}}
\renewcommand{\thesection}{S\arabic{section}}
\renewcommand{\thepage}{S\arabic{page}}

\onecolumngrid

\begin{center}
    {\bf Supplementary Material: Semi-fractal ground state in a 1D system with long-range hopping}\\
    
   Murod S. Bahovadinov, Faridun N. Jalolov, Vladimir E. Kravtsov, Boris L. Altshuler, Georgy V. Shlyapnikov
  
\end{center}
  
   In this supplementary material, we show the developed perturbation theory in momentum space in detail. We present the full derivation of the characteristics of the ground state wavefunctions, their moments and fractal dimensions, and the corresponding energies and compare them to the exact diagonalization results for various regimes.

 \section{ Second-order correction to the wavefunction}
 
 In this section we show the derivation of the probability density functions of the wavefunction components in the momentum space. Consider the results of the second-order perturbation theory for the wavefunctions: 
\begin{equation}\label{SecondOrder}
    \ket{\Psi}  = \ket{0}+ \sum_{K\neq0} \frac{U_K}{\Delta_K}\ket{K}  +\sum_{K\neq0,P\neq0}\frac{U_{K-P}U_P}{\Delta_K\Delta_P}\ket{K} -\sum_{K\neq0}\frac{U_KU_0}{\Delta_K^2}\ket{K}, 
\end{equation}

where the energy gap is $\Delta_Q=E_0-E_Q$, and the matrix elements of the disorder term $H_D$ are given by
\begin{equation}
    U_{k-k'}=\bra{k}H_D\ket{k'}=\frac{1}{L}\sum_{m}e^{i(k-k^\prime)m}e_m.
\end{equation}
The Fourier components of the uniformly distributed disorder variables $e_m\sim {\cal{U}}(-W,W)$ satisfy
\begin{equation}
\mathbb{E}[U_q]=0, \qquad 
\mathbb{E}[U_q U_{q'}^*]=s^2 \delta_{q,q'}, \qquad
\mathbb{E}[U_q U_{q'}]=s^2 \delta_{q,-q'},
\end{equation}
with $s^2 = \frac{W^2}{3L}$.

We note that the wavefunction given by (\ref{SecondOrder}) is not normalized. The third term at $P=K$ cancels the fourth term and one is left with the remaining sum:
 \begin{equation}
       \ket{\Psi}  = \ket 0+ \sum_{K\neq0}[\Psi^{(1)}(K)+\Psi^{(2)}(K) ]\ket{K},
 \end{equation}
 where 
  \begin{equation}
     \Psi^{(1)}(K)= \frac{U_{K}}{\Delta_K},
 \end{equation}
 and
 \begin{equation}
     \Psi^{(2)}(K)=\sum_{ P\neq0,P\neq K}\frac{U_{K-P}U_P}{\Delta_K\Delta_P}.
 \end{equation}
 \paragraph{Characteristic functions for the first-order term components.}
We first consider the real part of $ \Psi^{(1)}(K)$:
 
\begin{equation}\label{FirstOrder}
    \mathbf{Re} \Psi^{(1)}  =  \frac{1}{L} \frac{1}{\Delta_K}\sum_{m=1}^{L}\cos(Km)\epsilon_m= \frac{1}{L} \frac{S_1}{\Delta_K}
\end{equation}
 For the random variable $S_1$ one has the following characteristic function:
\begin{equation}
\Phi_{S_1}(t)=\prod_{m=1}^{L}\operatorname{sinc}[W\cos(Km)t].
\end{equation}
For large $\log(L)\gg1$, one can show that 
\begin{equation}
\Phi_{S_1}\!\left(t\right)
\xrightarrow[ ]{}
\exp\!\left(-\frac{W^{2}t^{2}}{12L}\right).
\end{equation}

Thus $S_1$ approaches a Gaussian random variable.
The same applies for the imaginary component.
Thus, the first-order term represents complex Gaussian variable for a given $K$, with the characteristic function (for the real and  imaginary parts):
\begin{equation}\label{FO}
    \Phi_{1}(K,t)= \exp\!\left(-\frac{\sigma ^2_1(K)t^2}{2}\right),
\end{equation}
where 
\begin{equation}\label{gamma1}
         \sigma^2_1(K) =\frac{W^2}{6L}\frac{1}{\Delta_K^2}.
\end{equation}
 \paragraph{Characteristic functions for the second-order term components.}

We next analyze the second-order term. For the pair of modes entering the second-order term,
\begin{equation}
\mathbb{E}[U_{K-P} U_P^*] = s^2 \delta_{K-P,P},
\qquad
\mathbb{E}[U_{K-P} U_P] = s^2 \delta_{K-P,-P}.
\end{equation}

Thus, for generic $K \neq 0$, both correlators vanish except at the resonant points satisfying
\begin{equation}
2P \equiv K \pmod{2\pi}.
\end{equation}

Away from these isolated momenta, $U_{K-P}$ and $U_P$ are uncorrelated (and, in the Gaussian limit, independent).
It should be emphasized that there are only $O(1)$ of the isolated terms with finite covariance, whereas for the other $O(L)$ terms the covariance vanishes. Thus, one can next assume the real and imaginary parts of all $X_P = \frac{U_{K-P} U_P}{\Delta_K \Delta_P}$ are then distributed according to a symmetric
double-sided Laplace law,
\begin{equation}
f_{\mathbf{Re} X_P}(x) = \frac{1}{2b_P} e^{-|x|/b_P},
\qquad
f_{\mathbf{Im} X_P}(y) = \frac{1}{2b_P} e^{-|y|/b_P},
\end{equation}
with the scale parameter
\begin{equation}
b_P = \frac{s^2}{2|\Delta_K \Delta_P|}.
\end{equation}

For the asymptotic distribution function of $\Psi^{(2)}(K)$ the CLT implies complex Gaussian distribution with the following characteristic functions for  both components: 
\begin{equation}\label{SOAResult}
    \Phi_2(t) = \exp\!\left(-\frac{\sigma^2_2(K)  t^2}{2} \right),
\end{equation}
 with the parameter 
  
 \begin{equation}
     \sigma^2_2(K) 
=
\frac{W^4}{18L^2\Delta_K^2}
\left[
\sum_{\substack{P\neq 0 \\ P\neq K}}\frac{1}{\Delta_P^2}
+
\sum_{\substack{P\neq 0 \\ P\neq K}}\frac{1}{\Delta_P\,\Delta_{K-P}}
\right].
 \end{equation} 

 \paragraph{ Covariance between first- and second-order contributions.}

We next consider covariance between the first- and second-order amplitudes at a fixed momentum $K\neq 0$. We note that both $\Psi^{(1)}(K)$ and $\Psi^{(2)}(K)$ are centered random variables. 
Thus, their covariance is
\begin{equation}
\mathrm{Cov}\big(\Psi^{(1)}(K),\Psi^{(2)}(K)^*\big)
=
\mathbb{E}\big[\Psi^{(1)}(K) \Psi^{(2)}(K)^*\big].
\end{equation}
Substituting the corrections one has
\begin{equation}
\mathbb{E}\big[\Psi^{(1)}(K) \Psi^{(2)}(K)^*\big]
=
\frac{1}{ \Delta_K ^2}
\sum_{P\neq 0,P\neq K}
\frac{1}{\Delta_P }
\mathbb{E}\big[U_K\, U_{K-P}^* U_P^*\big].
\end{equation}

Since the disorder Fourier modes are Gaussian with zero mean, all third-order moments vanish,
\begin{equation}
\mathbb{E}[U_a U_b U_c]=0,
\end{equation}
and therefore
\begin{equation}
\mathrm{Cov}\big(\Psi^{(1)}(K),\Psi^{(2)}(K)^*\big)=0.
\end{equation}
An analogous argument shows that
\begin{equation}
\mathrm{Cov}\big(\Psi^{(1)}(K),\Psi^{(2)}(K)\big)=0,
\end{equation}
so that all cross-correlations between first- and second-order terms vanish.
Thus, the first- and second-order contributions are uncorrelated and asymptotically independent in the large-$L$ limit. 
This allows one to write the following characteristic function for the unnormalized wavefunction (\ref{SecondOrder}) components:
\begin{equation}\label{FinalPhi}
    \Phi(K,t)=\Phi_1(K,t)\Phi_2(K,t)=\exp\!\left( -\frac{t^2}{4}\Sigma_{tot}^2(K) \right),
\end{equation}
where $\Sigma_{tot}^2(K)=2(\sigma^2_1(K)+\sigma^2_2(K))=\Sigma_1^2(K)+\Sigma_2^2(K).$

It is straightforward to show that $I_K=|\psi_{un}(K)|^2$ is exponentially distributed,

\begin{equation}
f(I_K)
=
\frac{1}{ \Sigma_{\mathrm{tot}}^2(K)}
\exp\!\left(-\frac{I_K}{ \Sigma_{\mathrm{tot}}^2(K)}\right),
\qquad I_K \ge 0.
\end{equation}
\paragraph{Large-system approximation for the normalized intensities.}

Since the unnormalized intensities satisfy
\begin{equation}
I_0=1,
\qquad
I_K \sim \mathrm{Exp}\!\big(\Sigma_{\mathrm{tot}}(K)^{-2}\big),
\qquad K\neq 0,
\end{equation}
the mean intensity for a given $K\neq0$ is $\mathbb{E}[I_K]=\Sigma_{\mathrm{tot}}^2(K).$
 
We define the normalized intensity as
\begin{equation}
\rho_K=|\psi(K)|^2=\frac{I_K}{\mathcal N},
\qquad
\mathcal N =1+\sum_{q\neq 0} I_q.
\end{equation}

In the large-system and weak disorder regime we approximate
\begin{equation}
\mathcal N \approx \mathcal S,
\qquad
\mathcal S =1+\sum_{q\neq 0}\Sigma_{\mathrm{tot}}^2(q)=1+\mathcal M.
\end{equation}
Hence, for $K\neq 0$, we have $\rho_K \approx \frac{I_K}{\mathcal S}.$
Since $I_K$ is exponentially distributed, $\rho_K$ is exponentially distributed random variable as well. 

Therefore,
\begin{equation}
\mathbb{E}[\rho_K]
\approx
\frac{\Sigma_{\mathrm{tot}}^2(K)}{\mathcal S},
\qquad
\operatorname{Var}(\rho_K)
\approx
\left(\frac{\Sigma_{\mathrm{tot}}^2(K)}{\mathcal S}\right)^2,
\qquad K\neq 0.
\end{equation}

For the zero-momentum component we have $\rho_0=\frac{1}{1+\sum_{k\neq 0} I_k}
\approx
\frac{1}{\mathcal S}$,
so that $\rho_0$ is approximately deterministic.

This approximation is valid provided
\begin{equation}
\frac{\sqrt{\sum_{k\neq 0}\Sigma_{\mathrm{tot}}^4(k)}}
{1+\sum_{k\neq 0}\Sigma_{\mathrm{tot}}^2(k)}
\ll 1,
\end{equation}
i.e. when the total norm is sharply concentrated around its mean.
 \section{ Comparison with exact diagonalization results}
 
\begin{figure}[t]
\includegraphics[width=\columnwidth]{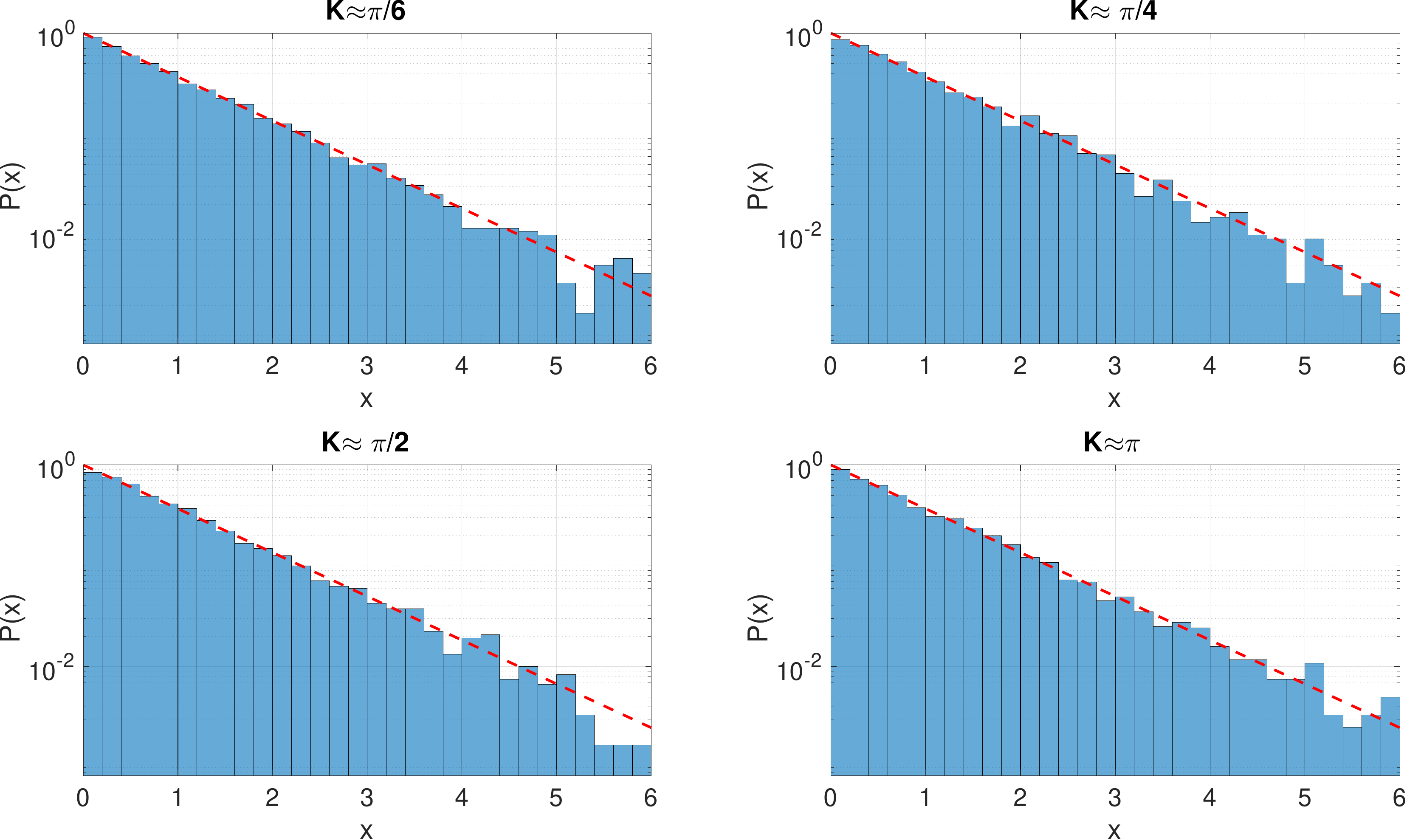}
\caption{Typical histograms for the reduced quantity $x$ obtained from ED calculations for $W=0.2$ at various $K$-points.  }
\label{fig:Fig_PDF}
\end{figure}

\begin{figure}[t]
\includegraphics[width=\columnwidth]{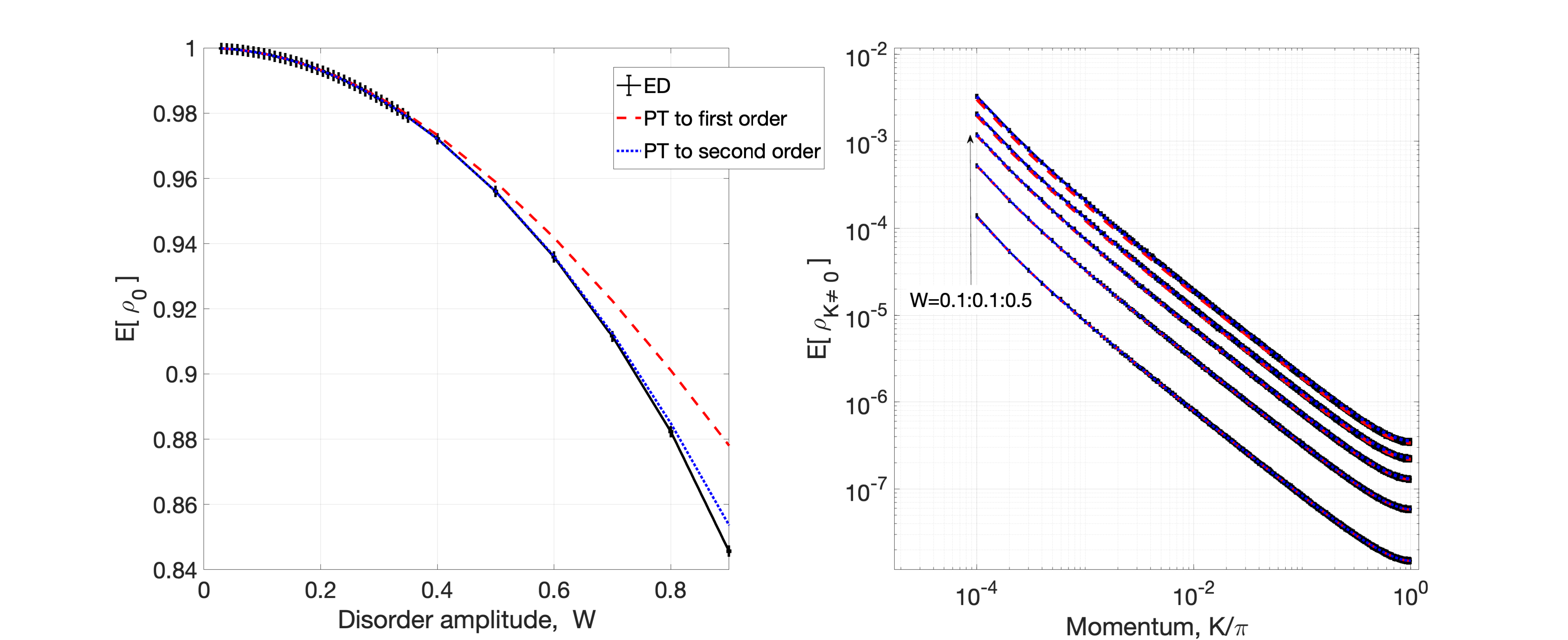}
\caption{
Comparison of the ED and PT results for the mean normalized intensity at zero momentum  $\mathbb{E}[\rho_0]$ (left panel) and the same for the tails shown on log-log scale (right panel).  The value of $a$ is $1.48$.
}
\label{fig:MeanRho}
\end{figure}

We now compare numerical ED results with the results of the perturbation theory (PT) presented in the previous section. 
We consider $a=1.48$ and $L=20000$, and the disorder realization number in ED is $N_D=6000$.
\paragraph{ Probability distribution function for $ \rho_k$} 
To test the predicted exponential form of the normalized intensities, we rescale the numerical data by their mean, $x=\rho_K/\mathbb{E[\rho_K]}$, and compare the resulting histograms to the universal law $P(x)=
\exp\!\left(-x\right)$. As shown in Fig.~\ref{fig:Fig_PDF}, the obtained histograms for the typical disorder amplitude $W=0.2$ nicely fit the exponential distribution at various parts of the Brillouin zone (BZ). Deviations are attributed to the finite statistics of the ED calculations. As an additional indicator, we next show the moment ratio $R_q=\mathbb{E[}\rho^q]/\mathbb{E[\rho]}^q$, which equals $\Gamma(1+q)$ for the exponential distribution for two values $q=1/2$ and $q=2$. The obtained results for $R_q$ indeed exhibit the constant values of $\Gamma(1+q)$ within the whole BZ, except at $K=0$. At $K=0$ the probability distribution is the inverse of a convolution of exponential distributions with non-identical parameters.
\begin{figure}[t]
\includegraphics[width=\columnwidth]{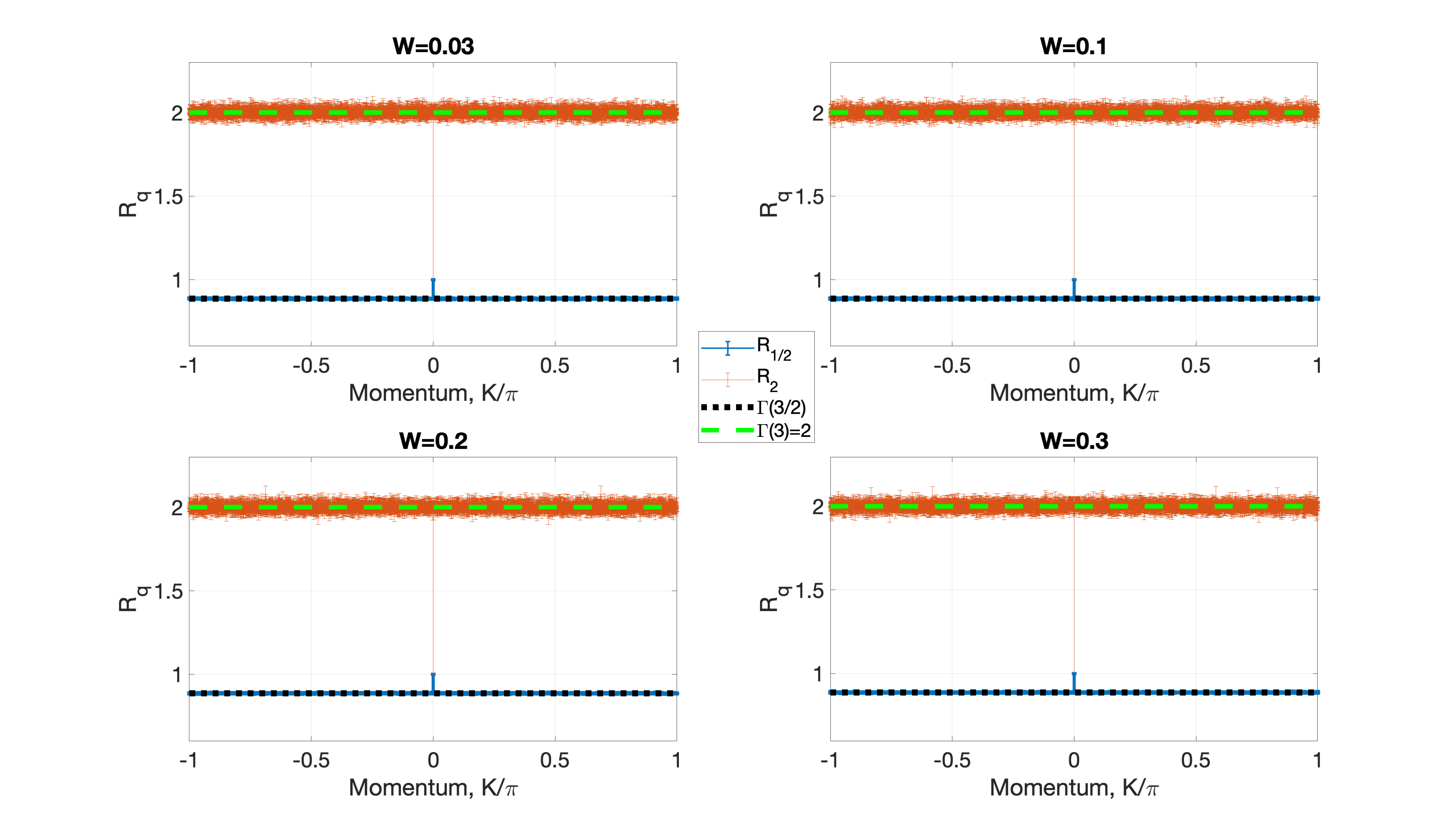}
\caption{Comparison of the PT and ED results for the moment ratio $R_q$ for several disorder amplitudes $W$ and $a = 1.48.$ }
\label{fig:r}
\end{figure}
\paragraph{ Results for $\mathbb{E}[\rho_k]$.} 
In Fig.~\ref{fig:MeanRho} we present the result of the numerical ED calculations and the PT results for $\mathbb{E}[\rho_k]$ at $K=0$ as a function of disorder amplitude $W$  (left panel) and at the tails for several disorder amplitudes (right panel). The mean value of the zero-momentum components calculated numerically is in good agreement with the PT results. The algebraic tails of the intensities are also correctly captured by the PT results. 

\begin{figure}[t]
\includegraphics[width= \columnwidth]{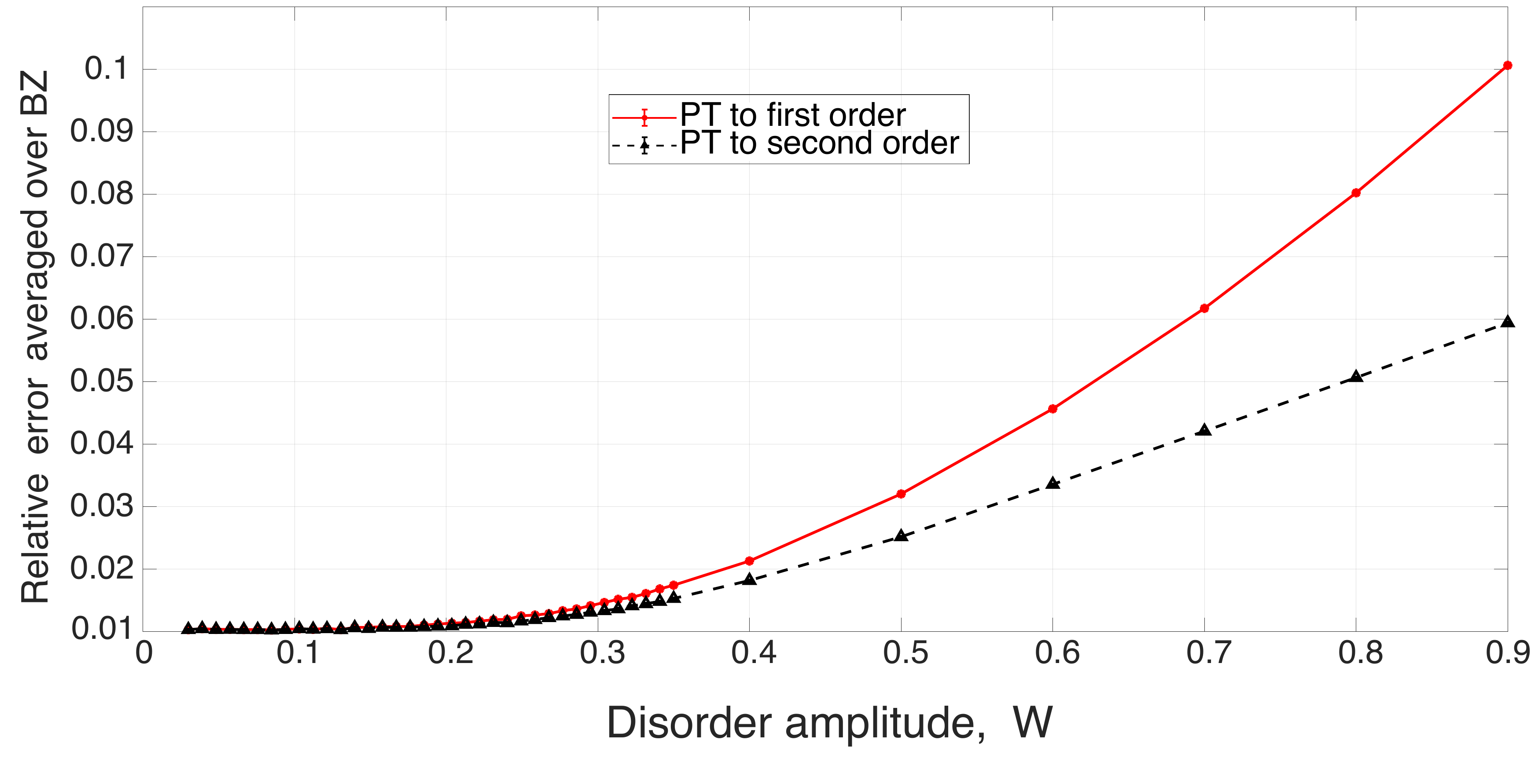}
\caption{ Brillouin-zone-averaged relative error $ \epsilon_\rho $  as a function of the disorder strength $W$. 
Error bars are smaller than the symbol size. }
\label{fig:error}
\end{figure}

To quantify the error, we next present the results for the relative error $\epsilon_\rho = \langle{{|\mathbb{E[\rho_K]_{PT}-\mathbb{E[\rho_K]_{ED}}}|}/{\mathbb{E[\rho_K]_{PT}}}}\rangle_K$ averaged over the BZ in Fig.~\ref{fig:error} for several values of the disorder amplitude $W$. The averaged relative error is also typically of order $10^{-2}$ for $W<0.2$, showing that perturbative analysis captures the mean normalized intensities with the accuracy of several percent.

\section{Moments of the wavefunctions and fractal dimensions}  

In this section we derive the asymptotic scaling of the wavefunction moments using the first-order PT results:
\begin{equation}
P_q(L)=\sum_k \mathbb{E}\!\left[\rho_k^q\right]=\sum_k\mathbb{E}[\frac{I_k^q}{\mathcal{S}(L)}]=\frac{1}{\mathcal{S}(L)}\sum_k\mathbb{E}[I_k^q]=\frac{1}{1+\mathcal{M}(L)}\sum_k\mathbb{E}[I_k^q],
\end{equation}
and the associated generalized fractal dimensions
\begin{equation}
D_q=\frac{\tau(q)}{q-1},
\qquad
P_q(L)\sim L^{-\tau(q)}.
\end{equation} 

\paragraph{Asymptotics of the energy denominators}

For the quantized momenta
\begin{equation}
k_m=\frac{2\pi m}{L},
\qquad
m=1,\dots,L,
\end{equation}
the small-momentum energy gap obeys
\begin{equation}
\Delta (k_m,a,L)
=
E(0,a,L)-E(k_m,a,L)
\sim
C_0(a)\left(\frac{2\pi m}{L}\right)^{a-1}+O(L^{-a-1}),
\end{equation}
with the prefactor
$C_0(a)=
-\Gamma(1-a)\sin\!\left(\frac{\pi a}{2}\right)$.

For convenience, we next define $\mu=2(a-1)$,
where $0<\mu<1$. We have
\begin{equation}
\Delta E(k_m,a,L)^2
\sim
C_0(a)^2(2\pi)^\mu L^{-\mu} m^\mu.
\end{equation}

\paragraph{Normalization factor.}

The perturbative normalization correction at first order is
\begin{equation}
\mathcal M (L)
=
\frac{W^2}{3L}
\sum_{m>0}\frac{1}{\Delta (k_m,a,L)^2}.
\end{equation}

Substituting the long-wavelength asymptotics yields
\begin{equation}
\mathcal M (L)
=
C_1(W,\mu)\,
L^{\mu-1}
\sum_{m=1}^{L}\frac{1}{m^\mu},
\end{equation}
with
$C_1(W,\mu)=
\frac{2W^2}{3\,C_0(a)^2(2\pi)^\mu}$.

The Euler-Maclaurin expansion for \(0<\mu<1\) leads to
\begin{equation}
\sum_{m=1}^{L}\frac{1}{m^\mu}
=
\frac{L^{1-\mu}}{1-\mu}
+\zeta(\mu)
+\frac12L^{-\mu}
+O(L^{-\mu-1}),
\end{equation}
whereas for \( \mu=1\) one has
\begin{equation}
\sum_{m=1}^{L}\frac{1}{m }
=\log(L)+\gamma_E+\frac{1}{2L}
+O(L^{-2}),
\end{equation}
 where $\gamma_E$ is the Euler-Mascheroni constant.
For \(0<\mu<1\) we obtain
\begin{equation}
\mathcal M (L)
=
C_1(W,\mu)
\left[
\frac{1}{1-\mu}
+\zeta(\mu)L^{\mu-1}
+\frac12L^{-1}
+O(L^{-2})
\right].
\end{equation}

Therefore, for $\mu<1$ corresponding to $a < \frac{3}{2}$, the normalization factor converges to a finite limit,
\begin{equation}
\mathcal M \xrightarrow[L\to\infty]{}
\mathcal M _\infty
=
\frac{C_1(W,\mu)}{1-\mu}.
\end{equation}
 Consequently, the normalized weight of the zero-momentum mode tends to
\begin{equation}
\rho_0(L)=\frac{1}{1+\mathcal M (L)}
\longrightarrow
\rho_0{(\infty)}
=
\frac{1}{1+\mathcal M_\infty}.
\end{equation}
For $\mu\geq1$ corresponding to $a\geq \frac{3}{2}$  the normalization factor $\mathcal{M}(L)$ diverges with increasing $L$. Hence, the $\rho_0(L)$ peak decreases and broadens.

\paragraph{Moments of non-zero modes.}

For \(m>0\), the perturbative normalized intensity is exponentially distributed with the mean
\begin{equation}
\mathbb{E}[\rho_m]
=
\frac{\Sigma^2_1(m)}{1+\mathcal M (L)},
\end{equation}
where
\begin{equation}
\Sigma^2_1(m)
=
\frac{W^2}{3L\,\Delta (k_m,a,L)^2}
\sim
C_1(W,\mu)L^{\mu-1}m^{-\mu}.
\end{equation}

Hence the moments obey
\begin{equation}
\mathbb{E}[\rho_m^q]
=
\Gamma(1+q)
\left(
\frac{\Sigma^2_1(m)}{1+\mathcal M(L) }
\right)^q.
\end{equation}

The total \(q\)-th moment reads
\begin{equation}
P_q(L)
=
\rho_0(L)^q
+
2\sum_{m=1}^{L}\mathbb{E}[\rho_m^q].
\end{equation}

Substituting the asymptotic form gives
\begin{equation}
P_q(L)
=
\rho_0(L)^q
+
C_2(q,W,\mu)\,
L^{q(\mu-1)}
\sum_{m=1}^{L}\frac{1}{m^{\mu q}},
\end{equation}
where
$C_2(q,W,\mu)=
2\Gamma(1+q)
\left(
\frac{C_1(W,\mu)}{1+\mathcal M_ \infty}
\right)^q.$


For \(0<q<1\), one has \(\mu q<1\), and therefore
\begin{equation}
\sum_{m=1}^{L}\frac{1}{m^{\mu q}}
\sim
\frac{L^{1-\mu q}}{1-\mu q}.
\end{equation}
Hence
\begin{equation}
P_q(L)\sim L^{1-q},
\qquad
0<q<1.
\end{equation}

It follows that
\begin{equation}
\tau(q)=q-1,
\qquad
D_q=1,
\qquad 0<q<1.
\end{equation}

For \(q>1\), the zero-mode contribution dominates,
$P_q(L)\to \rho_0{(\infty)}^q$. This leads to
 \begin{equation}
 \tau(q)=0,
 \qquad
D_q=0,
\qquad q>1. 
 \end{equation}

\paragraph{Shannon entropy and $D_1$.}
\begin{figure}[t]
\includegraphics[width= \columnwidth]{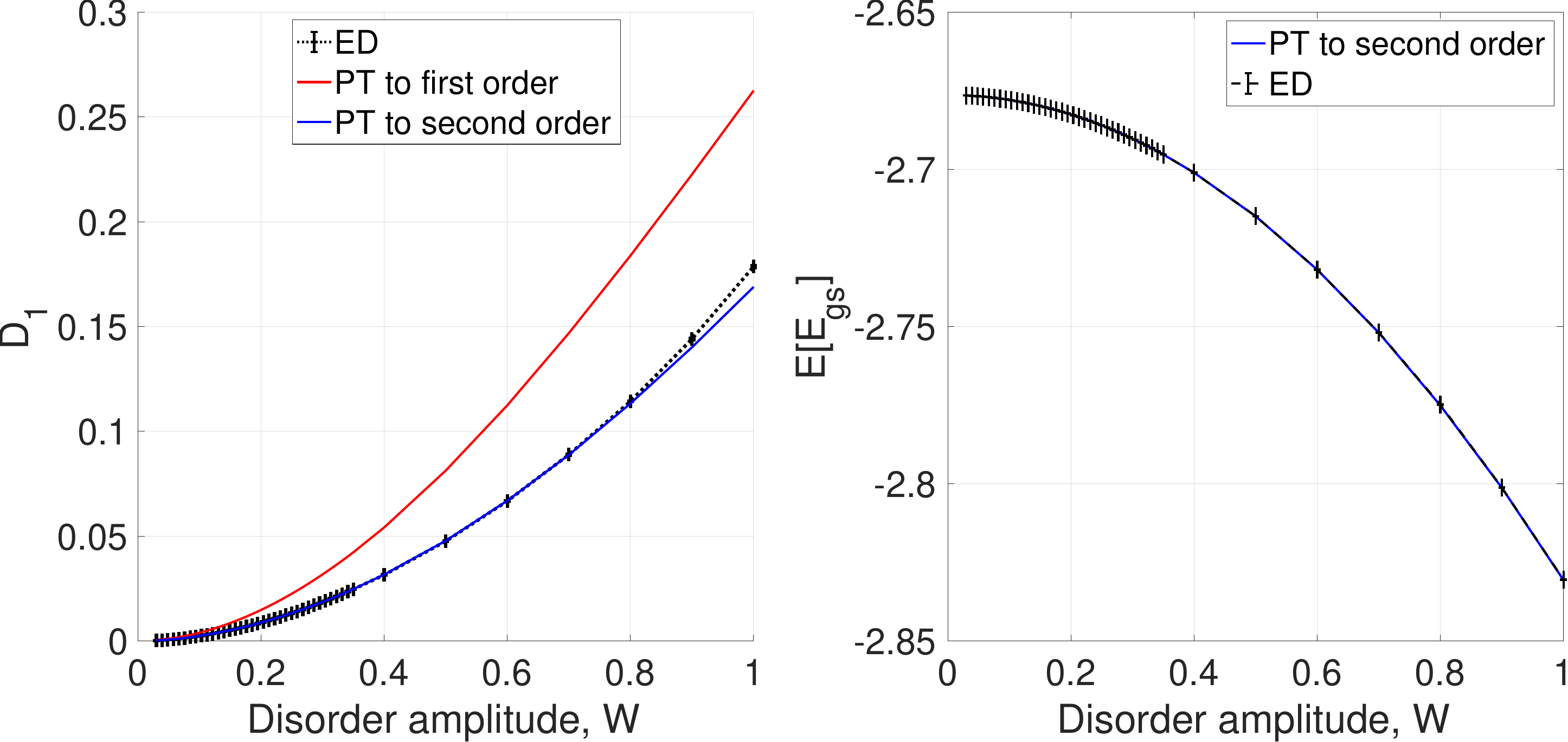}
\caption{ Comparison of the ED and PT results for the  fractal dimension $D_1$ (left panel) and for the averaged ground state energy $\mathbb{E}[E_{gs}]$ (right panel) as functions of disorder amplitudes $W$. Black dashed curve corresponds for the ED results. Error bars are smaller than the symbol size. }
\label{fig:D1andE}
\end{figure}

The case \(q=1\) is obtained from the Shannon entropy,
\begin{equation}
S_{\rm Sh}(L)
=
-\sum_K \mathbb{E}\!\left[\rho_K\ln\rho_K\right].
\end{equation}
Using
\begin{equation}
\mathbb{E}[x\log x]
=
\left.\frac{d}{dq}\mathbb{E}[x^q]\right|_{q=1},
\end{equation}
the Shannon entropy can be obtained from
\begin{equation}
S_{\rm Sh}(L)
=
-\left.\frac{dP_q(L)}{dq}\right|_{q=1}.
\end{equation}
For $K\neq0$, $\rho_K$ is exponentially distributed with the mean (first-order)
$  
\frac{\Sigma_1^2(K)}{1+\mathcal M(L)}$.
Therefore
$\mathbb{E}[\rho_K^q]
=
\Gamma(q+1)\frac{\Sigma_1^2(K)}{1+\mathcal M(L)}$,
and hence
\begin{equation}
\mathbb{E}[\rho_K\log\rho_K]
=
\frac{\Sigma_1^2(K)}{1+\mathcal M(L)}\left[\bar{\psi}(2)+\log\left(\frac{\Sigma_1^2(K)}{1+\mathcal M(L)} \right)\right],
\end{equation}
where $\bar{\psi}(x)$ is the digamma function.
For $1<a<3/2$, with $\mu \in(0,1)$, we have
\begin{equation}
\frac{\Sigma_1^2(m)}{1+\mathcal M(L)} \sim
\frac{ \,L^{\mu-1}m^{-\mu}}{1+\mathcal M_\infty}.
\end{equation}
This gives
\begin{equation}
S_{\rm Sh}(L)
=
\frac{\mathcal M_\infty}{1+\mathcal{M}_\infty}\log L+O(1).
\end{equation}
Thus one gets
$D_1=
\frac{\mathcal M_\infty}{1+\mathcal M_\infty}$. We compare this result with the ED result ($L=20000$) at $a=1.48$ in Fig.~\ref{fig:D1andE}.

\paragraph{Final Result}

Collecting all contributions one has
\begin{equation}
\boxed{
D_q=
\begin{cases}
1, & 0<q<1,\\[4pt]
\frac{\mathcal M _\infty}{1+\mathcal M _\infty}, & q=1,\\[10pt]
0, & q>1.
\end{cases}
}
\end{equation}

\begin{figure}[t]
\includegraphics[width= \columnwidth]{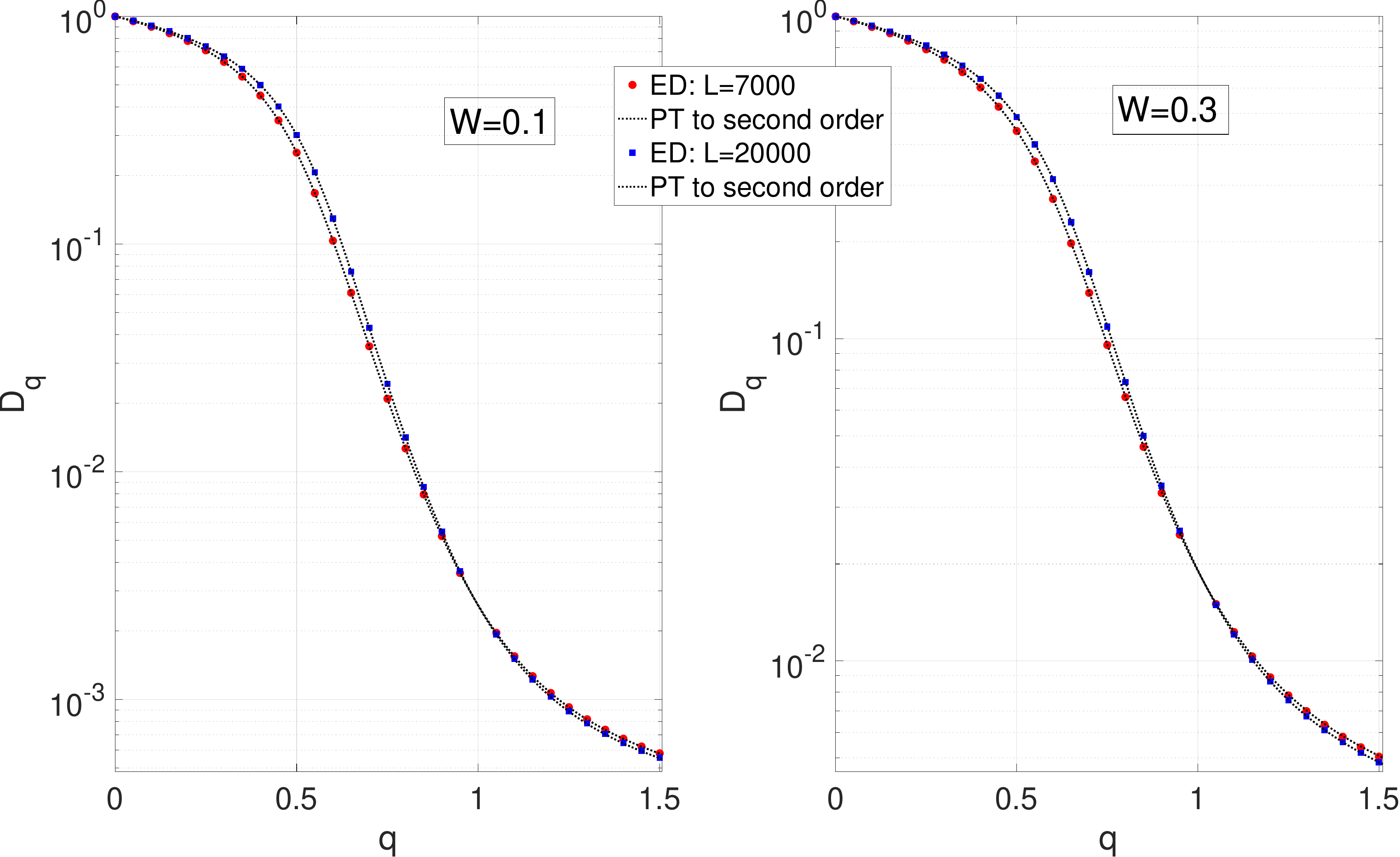}
\caption{ Comparison of the ED and PT results for the fractal dimensions $D_q$ as a function of $q$ for $W=0.1$ (left panel) and for $W=0.3$ (right panel). Black dashed curves corresponds for to second-order PT results. Error bars are smaller than the symbol size. The value of $a$ is $1.48$. }
\label{fig:DqVsQ}
\end{figure}

Thus, in the regime \(1<a<3/2\), the perturbative wavefunction exhibits multifractal spectrum with a discontinuity at \(q=1\). In Fig.~\ref{fig:DqVsQ} we compare the obtained PT results for $D_q$ against ED calculations for $L=7000$ and $L=20000$ at $a=1.48$ with $W=0.1$ (left panel) and $W=0.3$ (right panel).

\section{Results for the disorder-averaged ground-state energy}

We next consider the averaged ground state energy with the clean energy $E_0=-\zeta(a)+\zeta(a,L)$. For weak  disorder, the ground-state energy up to second order in perturbation theory is
\begin{equation}
E_{\mathrm{gs}}
=
E_0+\delta E^{(1)}+\delta E^{(2)},
\end{equation}
with
\begin{equation}
\delta E^{(1)} = U_0,
\qquad
\delta E^{(2)} = \sum_{K\neq 0}\frac{|U_K|^2}{\Delta_K},
\qquad
\Delta_K = E_0-E_K.
\end{equation}
 
The mean first-order correction vanishes immediately,
\begin{equation}
\mathbb{E}[\delta E^{(1)}]
=
\mathbb{E}[U_0]
=
0.
\end{equation}

For the second-order term we use $\mathbb{E}[|U_p|^2]=s^2$, which gives
\begin{equation}
\mathbb{E}[\delta E^{(2)}]
=
\sum_{K\neq 0}\frac{\mathbb{E}[|U_K|^2]}{\Delta_K}
=
 \sum_{K\neq 0}\frac{s^2 }{\Delta_K},
\end{equation}
with $s^2=\frac{W^2}{3L}$.
Therefore, the mean ground-state energy to second order is
\begin{equation}
\mathbb{E}[E_{\mathrm{gs}}]
=
E_0
+
 \sum_{K\neq 0}\frac{s^2}{\Delta_K}. 
\end{equation}

For the band minimum one has $E_0<E_K$ for all $p\neq 0$, so that
\begin{equation}
\Delta_P=E_0-E_K<0,
\end{equation}
and hence the second-order correction lowers the ground-state energy:
\begin{equation}
\mathbb{E}[E_{\mathrm{gs}}] < E_0.
\end{equation}
We present results for the mean ground state energies obtained from the ED and the PT calculations in the right panel of Fig.~\ref{fig:D1andE}.


\section{VALIDITY OF THE PERTURBATION THEORY}

To get qualitative idea on the range of validity of the developed perturbation theory, we next analyze the following ratio:
\begin{equation}
    R(L)=\frac{\sigma_2^2(K)}{\sigma_1^2(K)}=\frac{W^2}{3L}(S_1(K)+S_2(K)),
\end{equation}
where $S_1(K)=\sum_{P\ne0,K}\frac{1}{\Delta_P^2}$, and $S_2(K)=\sum_{P\neq0,K}\frac{1}{\Delta_P\Delta_{K-P}}$.
The growth of this ratio with the length of the system $L$ signals the breakdown of the perturbation theory.

 The analysis similar to the one in the previous sections leads to the following results:

\begin{equation}
    R(L)
    \sim
    \begin{cases}
     W^2  , & a<3/2,\\ 
     W^2\ln L, & a=3/2,\\ 
     W^2L^{2a-3} ,
        &  a>3/2.
    \end{cases}
    \label{eq:validity_of_PT}
\end{equation}
The result above implies that at $a=3/2$ the perturbation theory is still valid at small disorder. However the accuracy of the results decreases logarithmically with increasing $L$. For the case of $a>3/2$, on the contrary, the ratio diverges algebraically and one has the length-scale $L_*=W^{-\frac{2}{2a-3}}$ beyond which breakdown of the perturbation theory is observed.

 \section{SEMI-LOCALIZATION AT LARGE DISORDER}
\begin{figure}[t]
\includegraphics[width= \columnwidth]{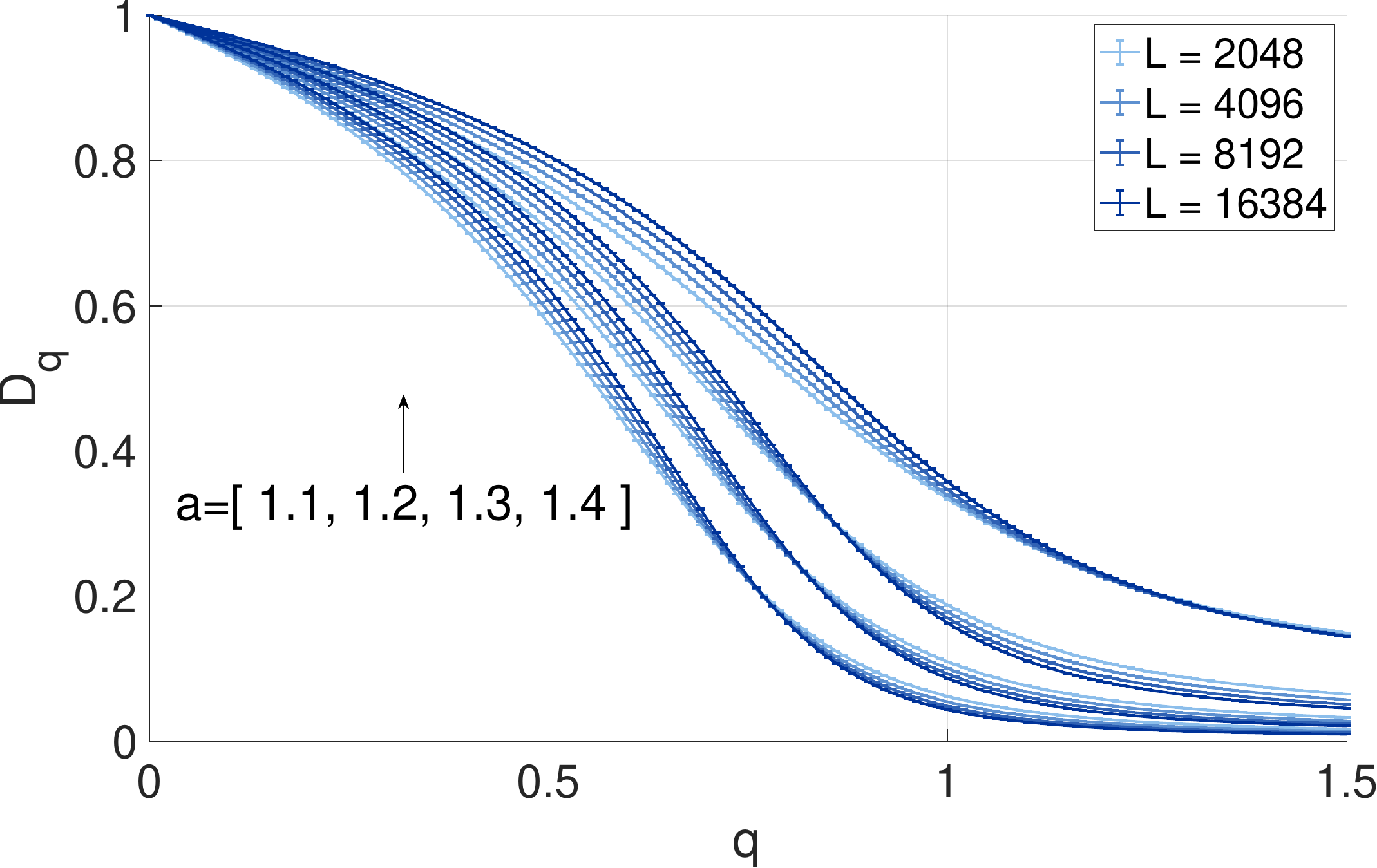}
\caption{ ED results for the fractal dimensions $D_q$ as a function of $q$ for $W=2$, various $a$ and $L$. The lower set of $4$ curves corresponds to $a=1.1$. From lightest to darkest, the blue curves correspond to system sizes \(L = 2048, 4096, 8192,\) and \(16384\) sites, respectively.   }
\label{fig:DqlargeDis}
\end{figure}

 We next present ED results for the fractal dimensions $D_q$ at fixed disorder strength $W=2$, various $a$ and $L$. The results are presented in Fig.~\ref{fig:DqlargeDis}. For $a<1.4$, the values $D_q$ increase with the system size for $q<q_*(L)$, whereas for $q>q_*(L)$ the quantity $D_q$ decreases with $L$ and $q_*(L\rightarrow\infty)=1$ is expected. For $a=1.4$ the critical point is $q_*(L)>1$ and it drifts towards larger values of $q$ with $L$, which signals that the semi-localized phase is already destroyed (the localized phase in the real space is realized). 

\section{MOMENTUM-SPACE LOCALIZATION FOR $a\leq 1$}

In the main text we focused on the regime $1<a<3/2$, where
the perturbative tails carry a finite fraction of the total norm in the
thermodynamic limit. We now show that this situation changes
qualitatively for $a\leq 1$. In this regime the total perturbative
weight of the tail, outside the zero-momentum mode, vanishes with increasing system
size. Consequently, the normalized ground-state wavefunction becomes
completely concentrated at $K=0$.

For $a<1$, the clean zero-momentum energy is nonextensive:
\begin{equation}
    E_0(a,L)
    \propto
    -\sum_{r=1}^{L-1}\frac{1}{r^a}
    =
    -\frac{1}{1-a}L^{1-a}
    +O(1),
    \label{eq:E0_alpha_less_one}
\end{equation}
 whereas for the gap one obtains
$    \Delta_m
    \propto
    L^{1-a}.$
 
Thus, the finite-momentum states become non-resonant with the
ground state, since the energy difference $E_K-E_0$ grows as $L^{1-a}$.

For the parameter $\mathcal M(L)$ one obtains 
\begin{equation}
    \mathcal M(L)\propto
    W^2L^{-2(1-a)}
    \longrightarrow0,
    \qquad a<1 .
    \label{eq:M_alpha_less_one_result}
\end{equation}

The mean intensity of an individual generic tail mode behaves as
\begin{equation}
    \left\langle I_K\right\rangle
    =
    \frac{W^2}{3L\Delta_K ^2}
    \sim W^2 
    L^{2a-3}.
    \label{eq:individual_tail_alpha_less_one}
\end{equation}
Although there are $O(L)$ such components, their accumulated weight
vanishes according to Eq.~\eqref{eq:M_alpha_less_one_result}.

Similarly, for $a=1$ one obtains the following result:
\begin{equation}
    \mathcal M(L)
    \sim
    \frac{W^2}{\ln^2L}
    \longrightarrow0,
    \qquad a=1 .
    \label{eq:M_alpha_one}
\end{equation}
Thus, $a=1$ is a marginal case in which the momentum-space tails
disappear logarithmically rather than algebraically.

Combining the results, the condensate depletion has the asymptotic
behavior
\begin{equation}
    1-\rho_0
    \sim
    \begin{cases}
    W^2L^{-2(1-a)}, & 0\leq a<1,\\[1mm]
    W^2/\ln^2L, & a=1,\\[1mm]
    \mathcal M_\infty/(1+\mathcal M_\infty),
        & 1<a<3/2.
    \end{cases}
    \label{eq:depletion_three_regimes}
\end{equation}
For $1<a<3/2$, the perturbative tails retain a finite total
weight, producing the semi-localized momentum-space state discussed in
the main text. For $a\leq1$, by contrast, their total weight
vanishes and only the $K=0$ component survives in the thermodynamic
limit. 
\section{Hidden SUSY structure for the problem of random walk in the presence of potential forces.}
The problem considered in Refs.\cite{KLY1, KLY2, KLY3} was formulated by the Fokker-Planck equation:
\begin{equation}\label{FP}
\frac{dP}{dt}={\cal L}P=-D\,\ \nabla\,(\nabla P + \beta\,(\nabla V)\,P),
\end{equation}
where $P$ is the probability density to find a particle at the point $r$ at a time $t$, $D$ is the diffusion coefficient, $V(r)$ is the random potential and $\beta$ is the inverse temperature. By making a substitution 
\begin{equation}
P=e^{-\beta\,V/2}\,\Psi,
\end{equation}
one reduces the Fokker-Planck equation to imaginary-time Shroedinger equation:
\begin{equation}
\frac{d\Psi}{dt}= \hat{H}\,\Psi=-D\,\nabla^{2}\,\Psi+ D\,U(r)\,\Psi,
\end{equation}
 where $U(r)$ has the celebrated Witten form \cite{SUSY}:
 \begin{equation}
U(r)=(\nabla\,W(r))^{2}-\nabla^{2}\,W(r),
 \end{equation}
 where
 \begin{equation}
W(r)=\frac{\beta}{2}\,V(r).
 \end{equation}
 With such a potential the Hamiltonian  $\hat{H}$ can be written as a product of two operators $A$ and $A^{\dagger}$:
 \begin{eqnarray}
A^{\dagger}&=&\nabla + \nabla\,W(r),\\
A&=&-\nabla + \nabla\,W(r),\\
\hat{H}&=&H_{-}=A A^{\dagger}.
 \end{eqnarray}
The partner Hamiltonian in the SUSY structure is $H_{+}=A^{\dagger}A$. As it was explained in the paper, one can define a {\it manifest chiral} Hamiltonian ${\cal H}$ in the extended Hilbert space:
\begin{equation}
{\cal H}=\left(\begin{matrix} 0 & A \cr A^{\dagger} & 0 \end{matrix} \right),
\end{equation}
which square contains both $H_{-}=\hat{H}$ and $H_{+}$:
\begin{equation}
{\cal H}^{2}=\left(\begin{matrix} AA^{\dagger} & 0 \cr 0 & A^{\dagger}A \end{matrix} \right).
\end{equation}
We see, therefore, that the Fokker-Planck equation Eq.(\ref{FP}) has a hidden SUSY structure similar to the one discussed in the main text for the ground state.

\end{document}